\documentclass[pdflatex, sn-nature]{sn-jnl}

\usepackage{graphicx}%
\usepackage{multirow}%
\usepackage{amsmath,amssymb,amsfonts}%
\usepackage{amsthm}%
\usepackage{mathrsfs}%
\usepackage[title]{appendix}%
\usepackage{xcolor}%
\usepackage{textcomp}%
\usepackage{manyfoot}%
\usepackage{booktabs}%
\usepackage{algorithm}%
\usepackage{algorithmicx}%
\usepackage{algpseudocode}%
\usepackage{listings}%
\usepackage[utf8]{inputenc}
\usepackage{graphicx}
\usepackage{amsmath}
\usepackage{amssymb}
\usepackage{pgfplotstable}
\usepackage{pgfplots}
\usepackage{chemfig}
\usepackage{natbib}

\theoremstyle{thmstyleone}%

\theoremstyle{thmstyletwo}%

\theoremstyle{thmstylethree}%

\newcommand{\Rjup}{\mbox{$R_{\rm Jup}$}} 
\newcommand{\Mjup}{\mbox{$M_{\rm Jup}$}}

\newcommand{\ltsimeq}{\raisebox{-0.6ex}{$\,\stackrel
        {\raisebox{-.2ex}{$\textstyle <$}}{\sim}\,$}}

\newcommand{\chhhh}{\mbox{${\rm CH_4}$}}
\newcommand{\hho}{\mbox{${\rm H_{2}O}$}} 
\newcommand{\coo}{\mbox{${\rm CO_{2}}$}} 
\newcommand{\nhhh}{\mbox{${\rm NH_{3}}$}} 
\newcommand{\phhh}{\mbox{${\rm PH_{3}}$}} 
\newcommand{\hhs}{\mbox{${\rm H_{2}S}$}}

\newcommand{\teff}{\mbox{$T_{\rm eff}$}}
\newcommand{\ldl}{\mbox{$\lambda/\Delta\lambda$}}
\newcommand{\logg}{\mbox{$\log{g}$}}

\makeatletter
\let\jnl@style=\rm
\def\ref@jnl#1{{\jnl@style#1}}

\def\aj{\ref@jnl{AJ}}                   
\def\actaa{\ref@jnl{Acta Astron.}}      
\def\araa{\ref@jnl{ARA\&A}}             
\def\apj{\ref@jnl{ApJ}}                 
\def\apjl{\ref@jnl{ApJ}}                
\def\apjs{\ref@jnl{ApJS}}               
\def\ao{\ref@jnl{Appl.~Opt.}}           
\def\apss{\ref@jnl{Ap\&SS}}             
\def\aap{\ref@jnl{A\&A}}                
\def\aapr{\ref@jnl{A\&A~Rev.}}          
\def\aaps{\ref@jnl{A\&AS}}              
\def\azh{\ref@jnl{AZh}}                 
\def\baas{\ref@jnl{BAAS}}               
\def\bac{\ref@jnl{Bull. astr. Inst. Czechosl.}}
\def\caa{\ref@jnl{Chinese Astron. Astrophys.}}
\def\cjaa{\ref@jnl{Chinese J. Astron. Astrophys.}}
\def\icarus{\ref@jnl{Icarus}}           
\def\jcap{\ref@jnl{J. Cosmology Astropart. Phys.}}
\def\jrasc{\ref@jnl{JRASC}}             
\def\memras{\ref@jnl{MmRAS}}            
\def\mnras{\ref@jnl{MNRAS}}             
\def\na{\ref@jnl{New A}}                
\def\nar{\ref@jnl{New A Rev.}}          
\def\pra{\ref@jnl{Phys.~Rev.~A}}        
\def\prb{\ref@jnl{Phys.~Rev.~B}}        
\def\prc{\ref@jnl{Phys.~Rev.~C}}        
\def\prd{\ref@jnl{Phys.~Rev.~D}}        
\def\pre{\ref@jnl{Phys.~Rev.~E}}        
\def\prl{\ref@jnl{Phys.~Rev.~Lett.}}    
\def\pasa{\ref@jnl{PASA}}               
\def\pasp{\ref@jnl{PASP}}               
\def\pasj{\ref@jnl{PASJ}}               
\def\psj{\ref@jnl{PSJ}}               
\def\rmxaa{\ref@jnl{Rev. Mexicana Astron. Astrofis.}}%
\def\qjras{\ref@jnl{QJRAS}}             
\def\skytel{\ref@jnl{S\&T}}             
\def\solphys{\ref@jnl{Sol.~Phys.}}      
\def\sovast{\ref@jnl{Soviet~Ast.}}      
\def\ssr{\ref@jnl{Space~Sci.~Rev.}}     
\def\zap{\ref@jnl{ZAp}}                 
\def\nat{\ref@jnl{Nature}}              
\def\iaucirc{\ref@jnl{IAU~Circ.}}       
\def\aplett{\ref@jnl{Astrophys.~Lett.}} 
\def\apspr{\ref@jnl{Astrophys.~Space~Phys.~Res.}}
\def\bain{\ref@jnl{Bull.~Astron.~Inst.~Netherlands}} 
\def\fcp{\ref@jnl{Fund.~Cosmic~Phys.}}  
\def\gca{\ref@jnl{Geochim.~Cosmochim.~Acta}}   
\def\grl{\ref@jnl{Geophys.~Res.~Lett.}} 
\def\jcp{\ref@jnl{J.~Chem.~Phys.}}      
\def\jgr{\ref@jnl{J.~Geophys.~Res.}}    
\def\jqsrt{\ref@jnl{J.~Quant.~Spec.~Radiat.~Transf.}}
\def\memsai{\ref@jnl{Mem.~Soc.~Astron.~Italiana}}
\def\nphysa{\ref@jnl{Nucl.~Phys.~A}}   
\def\physrep{\ref@jnl{Phys.~Rep.}}   
\def\physscr{\ref@jnl{Phys.~Scr}}   
\def\planss{\ref@jnl{Planet.~Space~Sci.}}   
\def\procspie{\ref@jnl{Proc.~SPIE}}   

\begin{document}
\title{Silicate Precursor Silane detected in Cold Low-Metallicity Brown Dwarf}
\author*[1,2]{\fnm{Jacqueline K.} \sur{Faherty}}\email{jfaherty@amnh.org}
\author[3]{\fnm{Aaron M.} \sur{Meisner}}
\author[4]{\fnm{Ben} \sur{Burningham}}
\author[5,26]{\fnm{Channon} \sur{Visscher}}
\author[6]{\fnm{Michael} \sur{Line}}
\author[1]{\fnm{Genaro} \sur{Su\'arez}}
\author[7,8]{\fnm{Jonathan} \sur{Gagn\'e}}
\author[1,2]{\fnm{Sherelyn} \sur{Alejandro Merchan}}
\author[1,2,9]{\fnm{Austin James} \sur{Rothermich}}
\author[13]{\fnm{Adam J.} \sur{Burgasser}}
\author[11]{\fnm{Adam C.} \sur{Schneider}}
\author[1]{\fnm{Dan} \sur{Caselden}}
\author[10]{\fnm{J. Davy} \sur{Kirkpatrick}}
\author[12]{\fnm{Marc Jason} \sur{Kuchner}}
\author[14]{\fnm{Daniella Carolina} \sur{Bardalez Gagliuffi}}
\author[15]{\fnm{Peter} \sur{Eisenhardt}}
\author[10]{\fnm{Christopher R.} \sur{Gelino}}
\author[16]{\fnm{Eileen C.} \sur{Gonzales}}
\author[10]{\fnm{Federico} \sur{Marocco}}
\author[17]{\fnm{Sandy} \sur{Leggett}}
\author[18,27]{\fnm{Nicolas} \sur{Lodieu}}
\author[19]{\fnm{Sarah L.} \sur{Casewell}}
\author[20]{\fnm{Pascal} \sur{Tremblin}}
\author[21]{\fnm{Michael} \sur{Cushing}}
\author[22]{\fnm{Maria Rosa} \sur{Zapatero Osorio}}
\author[18,27]{\fnm{V\'ictor J. S.} \sur{B\'ejar}}
\author[23]{\fnm{Bartosz} \sur{Gauza}}
\author[24]{\fnm{Edward} \sur{Wright}}
\author[25]{\fnm{Mark W.} \sur{Phillips}}
\author[18,27]{\fnm{Jun-Yan} \sur{Zhang}}
\author[18,27]{\fnm{Eduardo L.} \sur{Martin}}

\affil*[1]{\orgdiv{Department of Astrophysics}, \orgname{American Museum of Natural History}, \orgaddress{\street{79th street and CPW}, \city{New York}, \postcode{10023}, \state{NY}, \country{USA}}}

\affil[2]{\orgdiv{Department of Physics}, \orgname{The Graduate Center City University of New York}, \orgaddress{\city{New York}, \postcode{10016}, \state{NY}, \country{USA}}}

\affil[3]{\orgname{NSF National Optical-Infrared Astronomy Research Laboratory}, \orgaddress{\street{950 N. Cherry Ave.}, \city{Tucson}, \postcode{85719}, \state{AZ}, \country{USA}}}

\affil[4]{\orgdiv{Department of Physics, Astronomy and Mathematics}, \orgname{University of Hertfordshire}, \orgaddress{\street{AL10 9AB}, \city{Hatfield}, \country{United Kingdom}}}

\affil[5]{\orgdiv{Chemistry \& Planetary Sciences}, \orgname{Dordt University}, \orgaddress{\city{Sioux Center}, \postcode{51250}, \state{IA}, \country{USA}}}

\affil[6]{\orgdiv{School of Earth and Space Exploration}, \orgname{Arizona State University}, \city{Tempe}, \postcode{85287}, \state{AZ}, \country{USA}}

\affil[7]{\orgname{Plan\'etarium de Montr\'eal}, \orgaddress{\street{Espace pour la Vie, 4801 av. Pierre-de Coubertin}, \city{Montr\'eal}, \country{Canada}}}

\affil[8]{\orgdiv{D\'epartement de Physique}, \orgname{Trottier Institute for Research on Exoplanets, Universit\'e de Montr\'eal}, \orgaddress{\street{C.P.~6128 Succ. Centre-ville QC H3C~3J7 }, \city{Montr\'eal}, \country{Canada}}}

\affil[9]{\orgdiv{Department of Physics \& Astronomy}, \orgname{Hunter College}, \orgaddress{\street{695 Park Ave}, \city{New York}, \postcode{10065}, \state{NY}, \country{USA}}}

\affil[10]{\orgdiv{IPAC}, \orgname{Caltech}, \orgaddress{\street{1200 E. California Blvd.}, \city{Pasadena}, \postcode{91125}, \state{CA}, \country{USA}}}

\affil[11]{\orgdiv{Flagstaff Station}, \orgname{United States Naval Observatory}, \orgaddress{\street{ West Naval Observatory Rd.}, \city{Flagstaff}, \postcode{10391 }, \state{AZ}, \country{USA}}}

\affil[12]{\orgdiv{Science Mission Directorate}, \orgname{Mary W. Jackson NASA Headquarters }, \orgaddress{\street{300 E Street SW}, \city{Washington}, \postcode{20546}, \state{DC}, \country{USA}}}

\affil[13]{\orgdiv{Department of Astronomy \& Astrophysics}, \orgname{University of California, San Diego}, \orgaddress{\street{9500 Gilman Drive}, \city{La Jolla}, \postcode{92093}, \state{CA}, \country{USA}}}

\affil[14]{\orgdiv{Department of Physics \& Astronomy}, \orgname{Amherst College}, \orgaddress{\street{25 East Drive}, \city{Amherst}, \postcode{01002}, \state{MA}, \country{USA}}}

\affil[15]{\orgdiv{Jet Propulsion Laboratory}, \orgname{California Institute of Technology},\city{Pasadena}, \state{CA}}

\affil[16]{\orgdiv{Department of Physics \& Astronomy}, \orgname{San Francisco State University}, \orgaddress{\street{1600 Holloway Ave}, \city{San Francisco}, \postcode{94132}, \state{CA}, \country{USA}}}

\affil[17]{\orgdiv{NSF NOIRLab}, \orgname{Gemini Observatory}, \orgaddress{\street{670 N. A'ohoku Place}, \city{Hilo}, \postcode{96720}, \state{HI}, \country{USA}}}

\affil[18]{ \orgname{Instituto de Astrof\'isica de Canarias}, \orgaddress{\street{calle V\'ia L\'actea s/n}, \city{San Cristobal de La Laguna}, \postcode{38205}, \state{Tenerife}, \country{Spain}}}

\affil[19]{\orgdiv{School of Physics and Astronomy}, \orgname{University of Leicester}, \orgaddress{\street{University Road}, \city{Leicester}, \postcode{LE1 7RH}, \state{Leics}, \country{UK}}}

\affil[20]{ \orgname{Université Paris-Saclay, UVSQ, CNRS, CEA, Maison de la Simulation}, \orgaddress{\city{Gif-sur-Yvette}, \postcode{91191}, \country{France}}}

\affil[21]{\orgdiv{Ritter Astrophysical Research Center, Department of Physics and Astronomy}, \orgname{University of Toledo}, \orgaddress{\street{2801 W. Bancroft St.}, \city{Toledo}, \postcode{43606}, \state{OH}, \country{USA}}}

\affil[22]{\orgdiv{Centro de Astrobiología, CSIC-INTA}, \orgaddress{\street{Camino Bajo del Castillo s/n}, \city{Villanueva de la Cañada}, \postcode{28269}, \state{Madrid}, \country{Spain}}}

\affil[23]{\orgdiv{Janusz Gil Institute of Astronomy}, \orgname{University of Zielona Góra}, \orgaddress{\street{Lubuska 2}, \city{Zielona Góra}, \postcode{65-265}, \country{Poland}}}

\affil[24]{\orgdiv{Department of Physics and Astronomy}, \orgname{University of California - Los Angeles}, \orgaddress{PO Box 951547, \city{Los Angeles}, \postcode{90095-1547}, \state{CA}, \country{USA}}}

\affil[25]{\orgdiv{Institute for Astronomy}, \orgname{University of Edinburgh}, \orgaddress{\street{Royal Observatory, Blackford Hill, } \city{Edinburgh}, \postcode{EH9 3HJ}, \country{UK}}}

\affil[26]{\orgdiv{Center for Exoplanetary Systems}, \orgname{Space Science Institute}, \orgaddress{\city{Boulder}, \postcode{80301}, \state{CO}, \country{USA}}}

\affil[27]{ \orgname{Departamento de Astrof\'isica, Universidad de La Laguna}, \orgaddress{\street{Avenida Astrof\'isico Francisco S\'anchez, s/n}, \city{San Cristobal de La Laguna}, \postcode{38206}, \state{Tenerife}, \country{Spain}}}

\keywords{Brown dwarfs, Y dwarf stars}

\maketitle

\clearpage
\textbf{
Within 20 pc of the Sun there are currently 29 known cold brown dwarfs - sources with measured distances and an estimated effective temperature between that of Jupiter (170K) and $\sim$500K (\citep{Kirkpatrick24}).  These sources are almost all isolated and are the closest laboratories we have for detailed atmospheric studies of giant planets formed outside the solar system. Here we report JWST observations of one such source, WISEA J153429.75-104303.3 (W1534), which we confirm is a substellar mass member of the Galactic halo with a metallicity $<$0.01$\times$ solar. Its spectrum reveals methane (CH$_{4}$), water (H$_{2}$O), and silane (SiH$_{4}$) gas. Although SiH$_{4}$ is expected to serve as a key reservoir for the cloud-forming element Si in gas giant worlds, it eluded detection until now because it is removed from observable atmospheres by the formation of silicate clouds at depth. These condensates are favored with increasing metallicity, explaining why SiH$_4$ remains undetected on well studied, metal-rich solar system worlds like Jupiter and Saturn (\citep{Fegley94}). On the metal-poor world W1534, we detect a clear signature of SiH$_{4}$ centered at $\sim$4.55 $\mu$m with an abundance of 19$\pm$2 parts per billion (ppb). Our chemical modelling suggests that this SiH$_4$ abundance may be quenched at $\sim$ kilobar levels just above the silicate cloud layers, whereupon vertical atmospheric mixing can transport SiH$_4$ to the observable photosphere. The formation and detection of SiH$_{4}$ demonstrates key coupled relationships between composition, cloud formation, and atmospheric mixing in cold brown dwarf and planetary atmospheres.}

Our solar system's gas giants, Galactic brown dwarfs, and extrasolar planets define three different yet interconnected sub-fields of astrophysics. Nonetheless, they have overlapping temperatures that drive the chemical makeup of their atmospheres, so they have comparable observable properties. Jupiter has a photospheric temperature defined at the top of its visible cloud layers.  Ammonia (NH$_3$)  clouds can be found in the pressure range of 0.7 - 1.0 bar which corresponds to a temperature of $\sim$150 K (\citep{Fletcher2016}. Water (H$_2$O), methane (CH$_4$), NH$_3$ gases, and various other cloud layers can also be found throughout the upper atmosphere (\citep{Atreya2005}, \citep{Wong04}, \citep{Gapp24}).  Brown dwarfs span a range in temperature from $\sim$3000 K to $\sim$ 250 K, but the coldest examples have atmospheres similar in nature to that of Jupiter.  For example, the coldest known brown dwarf, WISE J085510.83-071442.5 (W0855; \citep{Luhman14}), has an effective temperature ({\teff}) $\sim$ 280 K (\citep{Luhman24}), and its atmosphere shows strong H$_2$O, CH$_4$, NH$_3$,  carbon dioxide (CO$_2$), and carbon monoxide (CO) absorption, along with hydrogen sulfide (H$_2$S) and trace amounts of phosphine (PH$_3$; \citep{Rowland24}).  The coldest directly-imaged exoplanet, Epsilon Indi Ab with {\teff} $\sim$ 275~K, has photometric measurements consistent with NH$_3$, CH$_{4}$, CO$_{2}$, and CO absorption and an unknown gas or condensate opacity source that suppresses 3-5 $\mu$m (\citep{Matthews2024}).

All of these objects are cool enough that multiple cloud layers can form at various condensation temperatures and pressures.  Importantly, silicate clouds form at depth in brown dwarfs and planets.  However, condensation efficiency in a low-temperature atmosphere is influenced by overall metallicity and dynamical mixing processes.
For instance, a reduced metal abundance can suppress the formation of condensates, as evidenced by theoretical models \cite{Lacy23} and supported by some observations \cite{Burgasser2007,Gonzales18}. 
Conversely, an enhanced metal abundance favors condensation, as evidenced by the absence of SiH$_{4}$ in the atmospheres
of Jupiter and Saturn \cite{Fegley94,Visscher2010}.   In this work, we report mid-infrared spectroscopic observations of the low metallicity brown dwarf W1534 where we have detected the silicate-bearing gas hydride SiH$_{4}$.

\begin{figure}[t]
\centering
\includegraphics[width=1\textwidth]{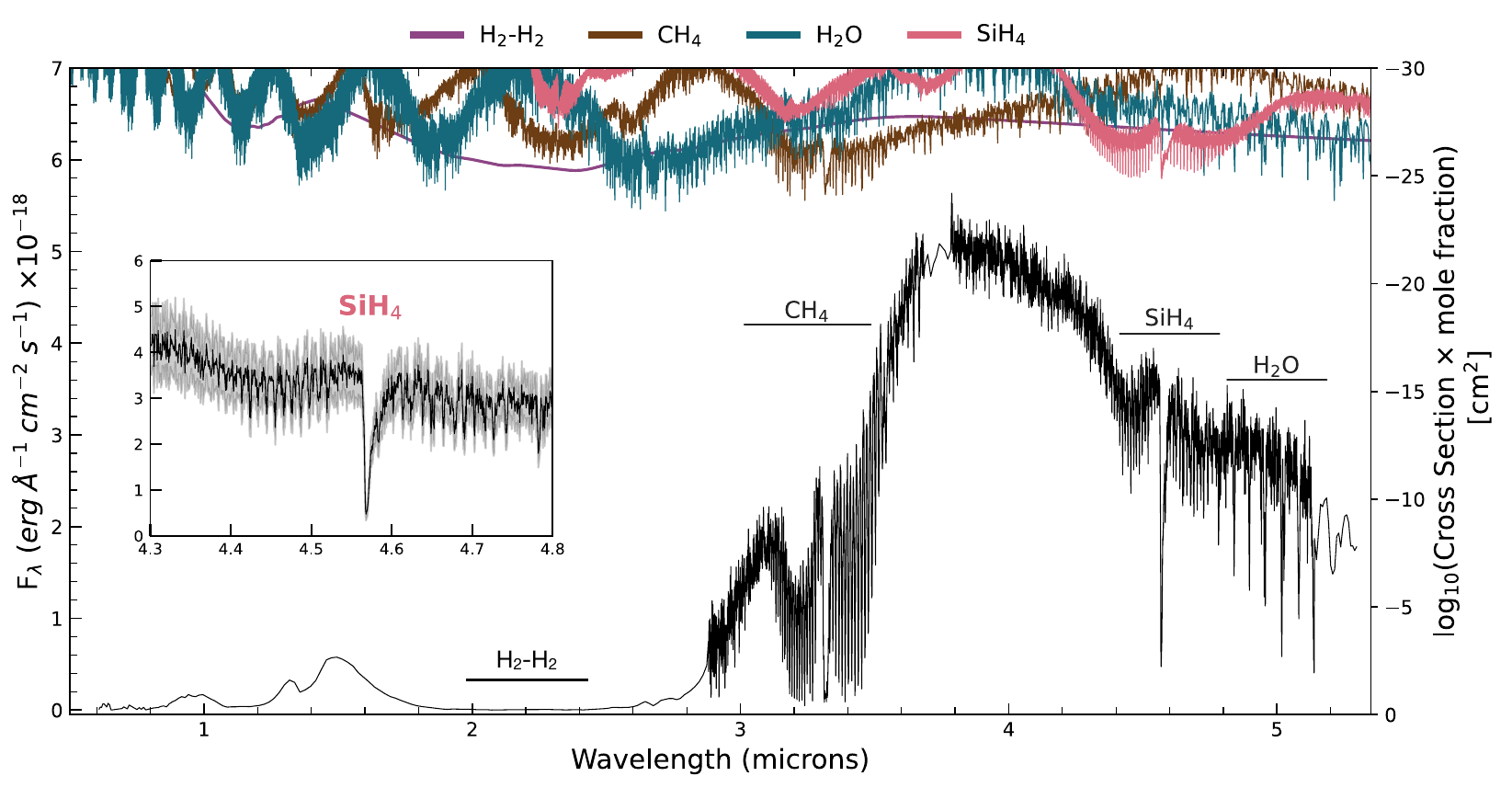} 
\caption{{\bf The distance-calibrated JWST NIRSpec prism and G395H portion of the spectral energy distribution for W1534.} 
Observed spectra are shown as a black line, with the G395H data between 2.9--5.1~$\mu$m having a higher resolution.
Cross sections multiplied by the retrieved mole fractions for H$_{2}$O, CH$_{4}$, SiH$_{4}$, and collision induced H$_{2}$ absorption (H$_{2}$-H$_{2}$) are shown as colored lines, and the regions in which they dominate
the absorption are labeled. The inset box shows the 4.4--4.7~$\mu$m region highlighting the silane feature with grey 1$\sigma$ uncertainties.}
\label{fig:spectra}
\end{figure}

W1534 was observed as part of the James Webb Space Telescope (JWST) Cycle 2 General Observer (GO) program 3558 (PI Meisner). JWST is one of NASA's flagship space-based observatories and specializes in examining the infrared portion of the sky. For GO program 2558, spectra were obtained with the Near-Infrared Spectrograph (NIRSpec) instrument in its low-resolution ({\ldl} $\approx$ 30--200) prism and high-resolution ({\ldl} $\approx$ 2000--4000) G395H modes. We also obtained mid-infrared photometry with the Mid-Infrared Instrument (MIRI) in the F1500W, F1800W, and F2100W filters centered at 15.0, 18.0, and 21.0 $\mu$m respectively.

We combined previously reported measurements of W1534, including its parallax and Wide-Field Infrared Explorer (WISE), Spitzer, Hubble Space Telescope (HST) and ground-based photometry, with the JWST spectra and photometry to create an absolute spectral energy distribution (SED) (see Extended Data Table 1).
Figure~\ref{fig:spectra} shows the NIRSpec portion of the SED. The only recognizable molecular absorption features in the spectrum are H$_{2}$O, CH$_{4}$, and the newly detected molecule SiH$_{4}$.  Shortward of 2.5~$\mu$m, W1534 looks like an extreme version of known low metallicity T dwarfs (\cite{Schneider2020,Lodieu22,Burgasser2024}),
with strongly suppressed 2~$\mu$m flux 
likely due to enhanced collision induced H$_{2}$ absorption (\citep{Linsky1969,Burgasser2003}). 
Longward of 2.5 $\mu$m, W1534 diverges from other cold substellar objects observed in the mid-infrared by Akari (\cite{Yamamura2010})
or JWST (\cite{Faherty24,Lew2024,Luhman24}) as there are no visible absorption features of ammonia (centered at $\sim$3~$\mu$m), carbon dioxide (centered at $\sim$4.3~$\mu$m) or carbon monoxide (centered at 4.8~$\mu$m).  

\begin{figure}[t]
\centering
\includegraphics[width=1.0\textwidth]{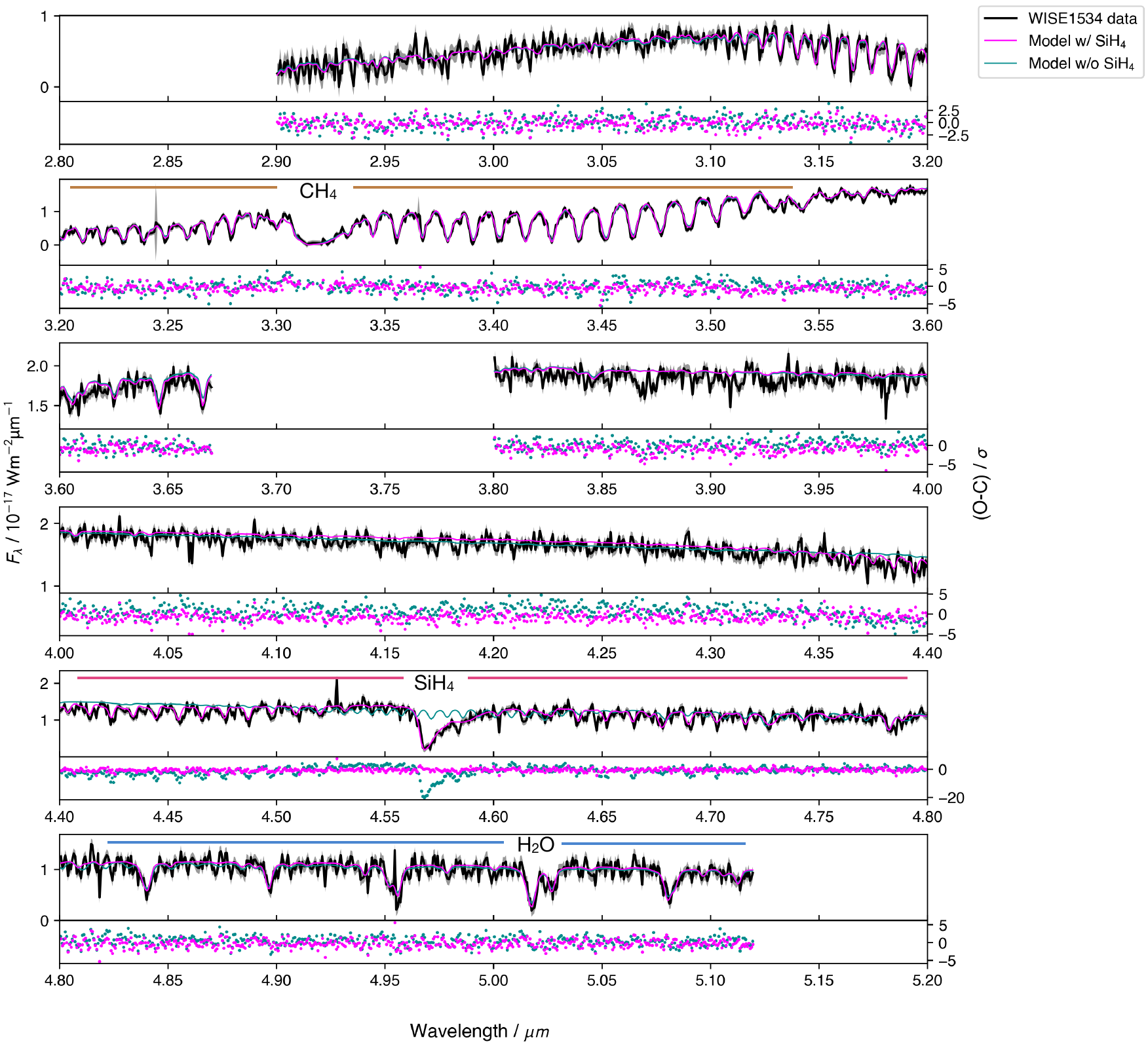} 
\caption{{\bf The JWST G395H spectrum for W1534 overlaid with the median likelihood retrieval model including (magenta) or excluding (teel) SiH$_{4}$.}  Data uncertainties are shaded in gray. The 67\% confidence intervals in the model posteriors are also shaded in grey, but are of comparable extent to the width of the plotted data. Below each of the five subpanels of the G395H spectrum are the residuals from the retrieval model. The major absorption bands of CH$_{4}$, SiH$_{4}$, and H$_{2}$O are labeled.}
\label{fig:w1534spec}
\end{figure}

Prior investigations of W1534 by \cite{Meisner20,Kirkpatrick21b} have argued that it is likely a very low-metallicity, low-temperature, substellar mass object. The JWST spectrum solidifies these claims. The cold nature of W1534 is demonstrated by its strong methane absorption at 3.3~$\mu$m, as we see this only in the coolest brown dwarfs (\cite{Beiler24}, \cite{Lew2024}, \citep{Faherty24}). The low metallicity of W1534 is demonstrated by the absence of CO and CO$_2$ bands, the suppresion of K band flux and, as discussed further, the detection of SiH$_{4}$. 

The kinematics of W1534 also support its low-metallicity nature.  Our retrieval analysis (see below and Methods section) determines a radial velocity of $-$116$\pm$4 km s$^{-1}$, which combined with published astrometry yields Galactic space velocity components of 
(U,V,W) = ($-$78$\pm$4,$-$192$\pm$15, $-$116$\pm$4) km s$^{-1}$, and a total velocity V$_{tot}$= 238 $\pm$ 14 km s$^{-1}$. Comparing to a collection of stars with full kinematics in the Gaia Catalog of Nearby Stars (\citep{GCNS21}, \citep{GaiaDR3}), W1534 sits squarely with halo objects (\cite{Bensby2014} see Methods section and Extended data Figure ~\ref{fig:Toomre}). 
Within 20~pc of the Sun there are only seven other sources with halo kinematics, none of which are substellar in nature (\cite{Kirkpatrick24}).

\begin{figure}[t]
\centering
\includegraphics[width=1.0\textwidth]{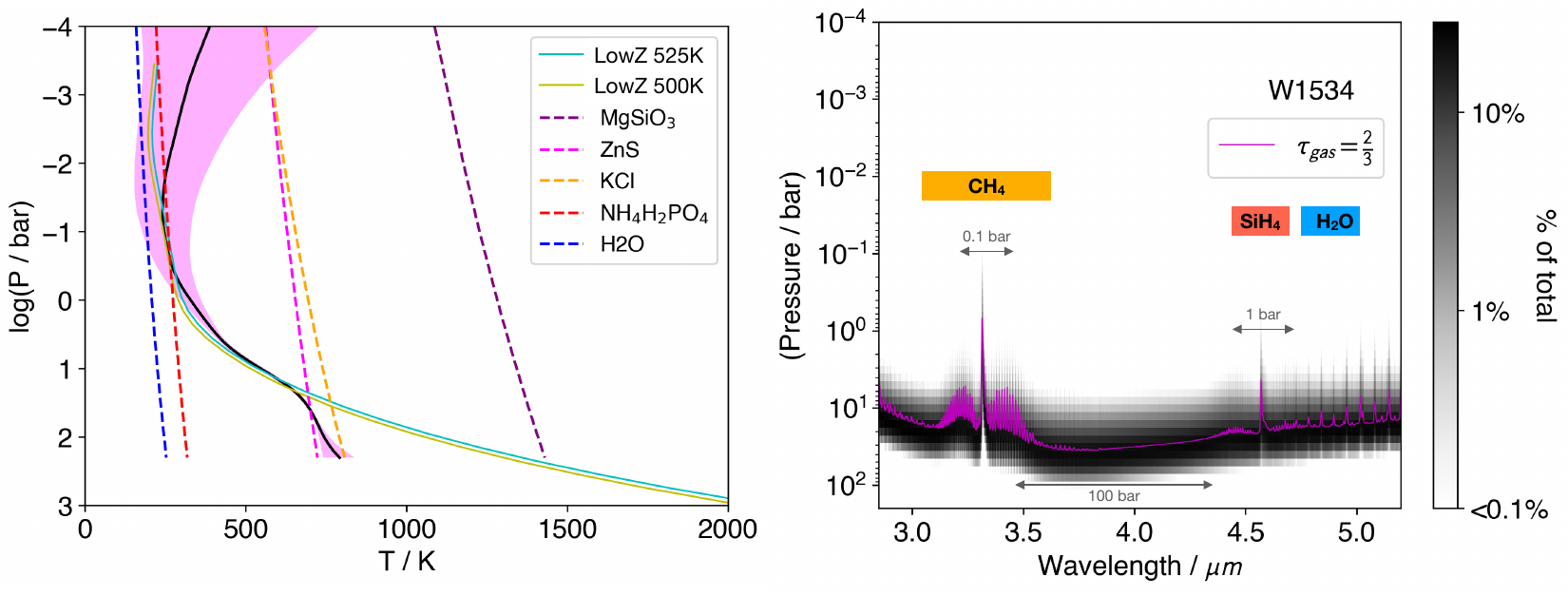} 
\caption{{\bf The best fit retrieved thermal profile and corresponding contribution function for W1534.} Left: The median posterior thermal profile for W1534 (black) with 67\% confidence interval (magenta shading).
Overlain are potential condensation curves for MgSiO$_3$ (purple dashed line), NH$_4$H$_2$PO$_4$ (red), H$_2$O (blue), KCl (orange), and ZnS (purple) as well as two self-consistent models from the LowZ grid (\citep{Meisner2021}). Comparison LowZ models have $[M/H]$=$-$2.25 with logg=4.9 and C/O=0.3.
Right:  The contribution function from the best fit retrieval (black line), with the $\tau_{gas}$=2/3 photosphere level highlighted in magenta. 
The locations of the primary absorption features from H$_{2}$O, CH$_{4}$, and SiH$_{4}$ are highlighted.} 
\label{fig:retrievedprofile}
\end{figure}

To better understand the molecular absorption features and quantify associated abundances, we conducted a retrieval modelling analysis of the NIRSpec G395H spectrum of W1534 using the {\it Brewster} framework (\citep{Burningham17,burningham2021}),
which has successfully retrieved the properties of numerous brown dwarfs (e.g. \citep{Gonzales20}, \citep{Vos22}, \citep{Calamari22}, \citep{Faherty24}).  For our baseline model we assumed a cloud-free observable photosphere. We performed the retrieval on just the G395H spectrum as this is where the higher resolution allows us to see detailed molecular absorption. The setup used was similar to that from \citep{Faherty24} which had the same input data from JWST and was an object of nearly the same temperature. Alongside continuum opacity sources described in detail in \citep{Burningham17,burningham2021}, the following gases were expected to have an impact in the wavelength range covered by our data: \hho, \chhhh, CO, \coo, \nhhh, \hhs, and \phhh. We included CH$_{3}$D and the isotopologues $^{12}$CO and $^{13}$CO as well.  Notably, the source for the absorption feature centered between 4.55 and 4.60 microns was not obvious at first glance given it had never been seen in a substellar mass object before.  Examining a suite of cross sections for molecular gases potentially present in gas giant planets, we realized that SiH$_{4}$ was a strong contender for the feature (see cross sections on Figure~\ref{fig:spectra}). Consequently, we tested two different retrieval set-ups: one with and one without SiH$_{4}$ gas.  Using a Bayesian Information Criterion (BIC) between the two tested models, the SiH$_{4}$ inclusive approach was strongly favored.  We note that for thoroughness we also ran a simple retrieval with a grey cloud but found no improvement on the fit. The abundances remained within 1$\sigma$ of the best-fit cloud-free model (see section 1.6.3 in the methods) therefore we leave cloud modeling for a future paper.

The results of the best fit retrieval are overplotted on Figure~\ref{fig:w1534spec}. This model affirms that the only detectable gases present in the atmosphere of this source are \hho, \chhhh, and SiH$_{4}$ (see Extended Data Table~\ref{tab:Abundances} in the Methods section and the corner plot in Extended Data Figure ~\ref{fig:w1534corner}). All other gases were undetected by the best retrieved model though within reason in a comparison to a low metallicity environment predicted from chemical equilibrium grid models (see Section 1.6.4 in methods and Extended data table~\ref{tab:Abundances} for upper limits).  The retrieval yielded several important fundamental parameters including a luminosity of log($L_{\rm bol}$/$L_{\odot}$) = $-$6.43$\pm$0.02, radius = 0.79$_{+0.07}^{-0.06}$ {\Rjup}; $T_{\rm eff}$ = 502$\pm$6 K; $\log g$ = 4.98$\pm$0.08 (in base units of cm/s$^2$), and mass = 19--30~{\Mjup}, placing W1534 amongst the least luminous and coldest substellar mass objects characterized to date (\citep{Beiler24}).  From the retrieved abundances of {\hho} and {\chhhh} we computed an overall atmospheric metallicity for W1534 of [M/H] = $-$2.22$\pm$0.05 and a carbon to oxygen (C/O) ratio of 0.26$^{+0.03}_{-0.02}$.  For comparison, FGK stars with [M/H] $<$ $-$1.0 have C/O ratios that plateau around a value of $\sim$ 0.2 (with scatter; see Extended data Figure~\ref{fig:CO}). The similarity in both bulk and relative abundances is evidence that W1534 formed similarly to metal-poor stars despite its much lower mass and temperature.

The detection of SiH$_{4}$ is an unexpected discovery from an observers perspective. Although Si is expected to be relatively abundant in the deep atmospheres of substellar objects (e.g. $\Sigma$Si $\sim$ 70 ppm in a solar-composition gas and $\sim$ 700 ppb in a [M/H] = $-$2.0 gas, using bulk element abundances from \citep{lodders2021}), Si-bearing species are effectively removed from the upper atmospheres of cool brown dwarfs and planets by the condensation of the Mg-silicates forsterite (Mg$_2$SiO$_4$) and enstatite (MgSiO$_3$) at depth (\citep{Fegley94,Visscher2010}). Noting an observational upper limit of 2.5 ppb on Jupiter (\citep{Treffers1978}) and 0.2 ppb on Saturn (\citep{Noll1991}), \citep{Fegley94} explored the possibility of transport-induced quenching of the SiH$_4\rightarrow$ silicate reaction pathway and demonstrated that SiH$_4$ quenches (i.e., where $t_{chem} = t_{mix}$) from atmospheric depths where it has already been depleted by silicate condensation. On these planets SiH$_4$ therefore remains below detection limits.

The left panel of Figure~\ref{fig:retrievedprofile} shows the retrieved temperature-pressure profile for W1534.  With increasing pressure, the retrieved profile is cooler than the LowZ forward model profile (\citep{Meisner2021}).  The ``kink" seen around 10 bar pressure where the temperature-pressure profile appears to become more isothermal has also been retrieved -- with differing levels of amplitudes -- in other substellar sources (see e.g. \citep{Faherty24}, \citep{Lew2024}, \citep{Kothari2024}).  It is yet to be seen if this is a consequence of an undetected and un-characterized cloud (e.g. NH$_{4}$H$_{2}$PO$_{4}$ cloud \citep{Morley18}) or  a misunderstood internal process (e.g. fingering convection \citep{Tremblin15}).  Either way, the contribution function in the right panel of Figure~\ref{fig:retrievedprofile} shows that the flux in W1534 emerges from almost 100 bar in pressure. As expected for a low-metallicity source, this is almost twice as deep as any other cold substellar mass object observed (e.g. \citep{Faherty24}, \citep{Kothari2024}, \citep{Lew2024}). 

Extrapolating the retrieved profile of W1534 to even deeper pressures (see Extended Data description and Extended Data Figure~\ref{fig:silane}), we find that SiH$_4$ can plausibly quench to abundances of $\sim 20$ ppb on W1534 for vertical diffusion coefficient values ($K_{zz}$) of $\sim$ $10^7-10^8$ cm$^2$ s$^{-1}$. The difference in SiH$_4$ quenching behavior on W1534 compared to cool giant planets such as Jupiter and Saturn can be attributed to two major factors. First, the very low metallicity ([M/H] $\sim$ $-$2.2) and thus highly reducing atmosphere of W1534 favors the formation of hydrides such as SiH$_4$ (\citep{Visscher2010}) relative to other Si-bearing gases (SiO, SiS), which expands the pressure and temperature region over which SiH$_4$ is the dominant Si-bearing gas and also lowers the temperature at which Si-bearing gases are removed by silicate condensation. Second, the higher gravity of W1534 ({\logg} $\sim 5.2$) compared to Jupiter ({\logg} $\sim 3.4$) yields a much shorter scale height and a shorter timescale ($t_{mix}$) for vertical mixing, which in turn increases the temperature at which SiH$_4$ oxidation to SiO quenches (i.e., $t_{chem} = t_{mix}$). As a result, SiH$_4$  can be mixed upward from depths where Si-bearing gases have not yet been completely depleted from the vapor phase by silicate cloud formation. The detection of SiH$_{4}$ in a substellar mass object 
therefore highlights the competing roles that temperature, pressure, composition, condensation, and vertical mixing play in determining the atmospheric chemistry of cool atmospheres, from the gas giants in our Solar System to brown dwarfs and exoplanets beyond. At present, no self-consistent forward model for substellar mass objects has included SiH$_{4}$ as an opacity source therefore this result motivates its inclusion moving forward especially for low metallicity objects. While W1534 is the only source with a spectral detection of SiH$_4$ to date, there are a number of JWST GO programs targeting low metallicity or unusual substellar mass objects where one might find the molecule.  Furthermore, the cross-section for SiH$_4$ indicates another strong absorption feature should appear between 10 - 12 $\mu$m. W1534 is slated for a MIRI Low Resolution Spectrometer (LRS) observation in the summer of 2025 which will cover this wavelength and confirm the 4.55$\mu$m detection.

\clearpage

\clearpage

\renewcommand{\tablename}{Extended Data Table}
\renewcommand\thetable{\arabic{table}}
\setcounter{table}{0}

\renewcommand{\figurename}{Extended Data Fig.}
\renewcommand\thefigure{\arabic{figure}}
\setcounter{figure}{0}

\clearpage
\section{Methods} \label{sec:methods}

\subsection{WISEA J153429.75-104303.3}\label{sec:sample}
\begin{figure}[t]
\centering
\includegraphics[width=1\textwidth]{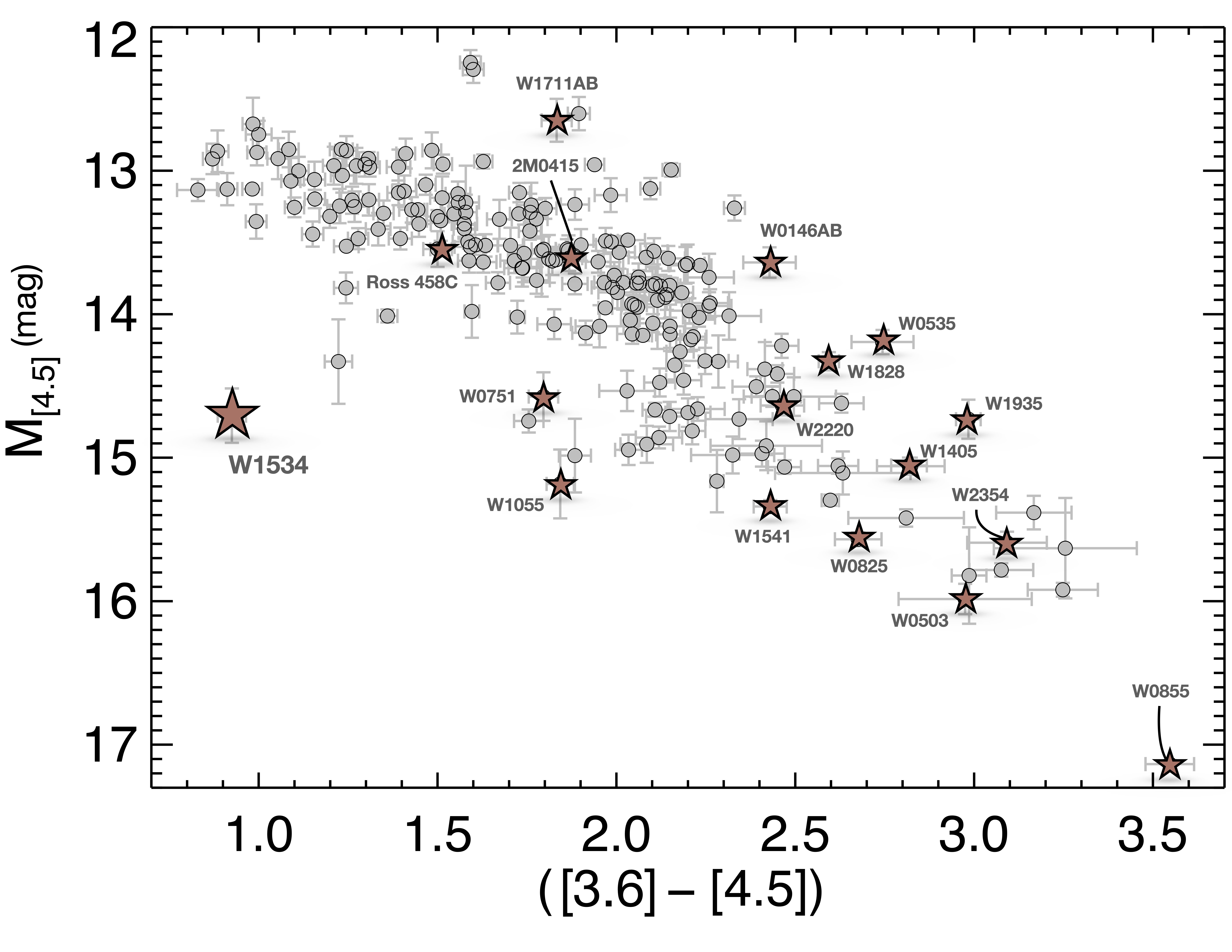} 
\caption{{\bf The Spitzer infrared color-magnitude diagram for cold brown dwarfs.}  The sources in JWST GO programs with comparable colors or magnitudes, as well as publicly available G395H or G395M spectra, are labeled and indicated by stars. The position of W1534 is the largest of the five-point stars which are labeled. Error bars indicate 1$\sigma$ uncertainties.}
\label{fig:CMD}
\end{figure}

WISEA J153429.75-104303.3 (W1534) was first discovered by citizen scientist Dan Caselden who used machine learning techniques on WISE data to try and recover cold, nearby substellar mass objects.   W1534 was first reported in \citep{Meisner20} and was initially noteworthy due to its blue Spitzer ([3.6] - [4.5]) color ($\sim$0.925 mag) and high proper motion ($\sim$2.7 arcsec yr$^{-1}$).  \citep{Kirkpatrick21b} followed it up with HST imaging and discovered it was at a nearby distance of 16.3$^{+1.4}_{-1.2}$ pc.  This distance marked it as an outlier on color magnitude diagrams (Extended Data Figure~\ref{fig:CMD}),
suggesting that W1534 was either extremely blue for its low luminosity or extremely faint for its temperature.
The goal of JWST Cycle 2 GO program 3558 (PI Meisner) was to target W1534 with NIRSpec and MIRI in order to provide a first spectroscopic assessment of its atmosphere, including temperature (estimated at $<$500~K)
and metallicity (estimated at [M/H] $\approx$ $-2$ \cite{Kirkpatrick21b}). Extended Data Figure~\ref{fig:CMD} shows the Spitzer color magnitude diagram for all cold brown dwarfs with parallaxes and highlights both the position of W1534 as well as targets that have similar JWST NIRSpec G395H observations.

\subsection{The Data} \label{sec:data}
W1534 was observed with the JWST Near-Infrared Spectrograph (NIRSpec \cite{Jakobsen2022}) in low-resolution prism and high-resolution G395H spectroscopic modes on 18 July 2024 (UT).
The source was acquired through the CLEAR filter using the wide aperture target acquisition (WATA) method, the SUB2048 subarray, and the NRSRAPID readout pattern. NIRSpec high resolution data were obtained using the F290LP filter, the G395H grating, the S200A1 slit, and the SUBS200A1 subarray which provides wavelength coverage of 2.87--5.14~$\mu$m with an average resolving power of $\sim$2700.  
NIRSpec low resolution data were obtained using the CLEAR filter, the prism grating, the S200A1 slit, and the SUBS200A1 subarray which
provides wavelength coverage of 0.6--5.3~$\mu$m with an average resolving power of $\sim$100.  
The G395H setup used 28 groups per integration, 18 integrations per exposure, and 3 dither positions for a total of 54 integrations in 2355.696 seconds of exposure time.  The prism setup used 35 groups per integration, 10 integrations per exposure, and 3 dithers for a total of 30 integrations in 6543.6 seconds of exposure time. Recorded time including overhead for the NIRSpec observations was 3.60 hours.   

Photometric imaging with the JWST Mid-Infrared Imager (MIRI \cite{Glasse2015}) was obtained 08 July 2024 (UT) with the F1500W (15~$\mu$m), F1800W (18~$\mu$m), and F2100W (21~$\mu$m) filters.  For each filter, the FASTR1 readout pattern was chosen with a 4-point dither pattern.  
The observing strategy employed 5 groups per integration for F1500W (56~s total exposure), 8 groups per integration for F1800W (89~s), and 16 groups per integration for F2100W (366~s).  Recorded time including overhead for the MIRI observations was 0.90 hours. 

W1534 has previously published photometry and astrometry which are included in Extended Data Table ~\ref{tab:target}.

\subsection{Data Reduction} \label{sec:datareduction}
NIRSpec prism and G395H spectra were reduced from uncalibrated data downloaded from the Barbara A.~Mikulski Archive for Space Telescopes (MAST) using the official JWST science calibration pipeline version 1.15.1 (\cite{Bushouse24}), based on the Calibration Reference Data System (CRDS \cite{Greenfield16}) context file jwst\_1276.pmap. The pipeline is comprised of three separate stages. 
Stage 1 performs detector-level corrections on the uncalibrated files (e.g. bias subtraction, dark subtraction, and cosmic-ray detection) and uses ramp fitting to generate count rate images for all exposures. Stage 2 calibrates these rate images by applying instrument-level and observing mode corrections. Stage 3 combines multiple calibrated exposures and extracts the spectrum from the combined image.

We ran the three pipeline stages with default parameters, but changed the resampling weight\_type parameter to ``ivm'' in Stages 2 and 3 to optimize the removal of noise spikes in the spectra. Flux uncertainties were automatically propagated through the pipeline and correspond to the combination in quadrature of the Poisson variance (\texttt{FLUX\_VAR\_POISSON}), read noise variance (\texttt{FLUX\_VAR\_RNOISE}), and 
flat variance (\texttt{FLUX\_VAR\_FLAT}).

The final calibrated NIRSpec prism spectrum spans the wavelength range 0.55--5.38~$\mu$m and has a median signal-to-noise ratio (SNR) of 240. The final calibrated NIRSpec G395H spectrum spans the wavelength range 2.86--5.14~$\mu$m, with a gap between the NRS1 and NRS2 NIRSpec detectors from 3.69 to 3.78~$\mu$m. The peak SNR for the G395H spectrum is 50 and the median SNR in each detector is 13 for NRS1 and 24 for NRS2. We note that the SiH$_{4}$ band centered around 4.55$\mu$m is present in both prism and G395H data as well as in each individual spectral dither.

For the MIRI data, the archival pipeline-based reduction detected the target in the F1500W filter but failed to do so in the F1800W and F2100W filters.  We therefore re-ran the version 1.15.1 pipeline on these data, altering the pipeline SNR threshold value -- which refers to the SNR above the background -- to 1.5 and 1 for the F1800W and F2100W filters, respectively. We choose the {\it aper total vegamag} column as our preferred magnitude and ensured that source fluxes remained consistent within uncertainties after changing the SNR value in a given filter.  In this manner, we recovered the target in each of the three filters and report the corresponding magnitudes in Extended Data Table~\ref{tab:target}.

We note that we did not encounter any issues with the JWST pipeline that appeared to compromise data quality. The pipeline products were deemed robust for our analysis.
\subsection{Radial Velocity} \label{sec:RV}
The radial velocity of W1534 is a fit parameter in the Brewster retrieval.  We used a uniform prior with a bounded range of $\pm$250 km s$^{-1}$, and note that the pipeline-reduced NIRSpec G395H spectrum is already corrected for barycentric motion (VELOSYS header value 25658.21 m s$^{-1}$). The corner plot in Extended Data Figure \ref{fig:w1534corner} shows a well-constrained retrieved value of $-$116$\pm$3~km~s$^{-1}$. As a conservative approach, we include a 
2.5~km~s$^{-1}$ systematic uncertainty in quadrature with this measurement to account for JWST pointing error and the under-sampled point spread function (see \citep{Gardner2006} and JWST user documentation on ``JWST Pointing Accuracy'').  Our final radial velocity with uncertainty is $-$116$\pm$4~km~s$^{-1}$.

\begin{figure}[t]
\centering
\includegraphics[width=1\textwidth]{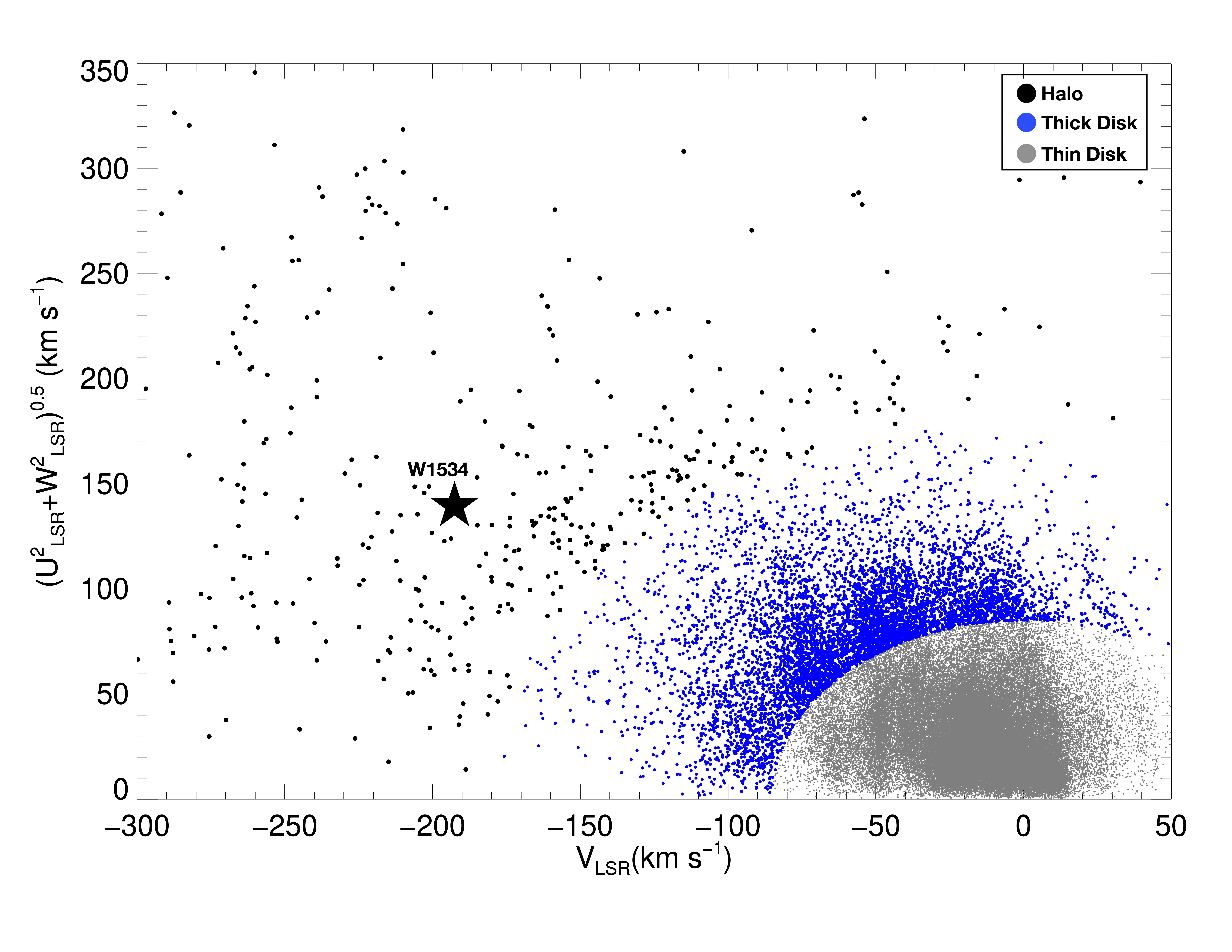} 
\caption{{\bf Toomre diagram of UVW space motions of nearby stars and brown dwarfs highlighting W1534.}  Points indicate sources with 3D kinematics from the 100~pc Gaia Catalog of Nearby stars (GCNS) containing radial velocities from Gaia DR3.  Thin disk (grey), thick disk (blue), and halo (black) sources are indicated by color; the location of W1534 is indicated by the large star.}
\label{fig:Toomre}
\end{figure}

\begin{figure}
\centering
\includegraphics[width=1.0\textwidth]{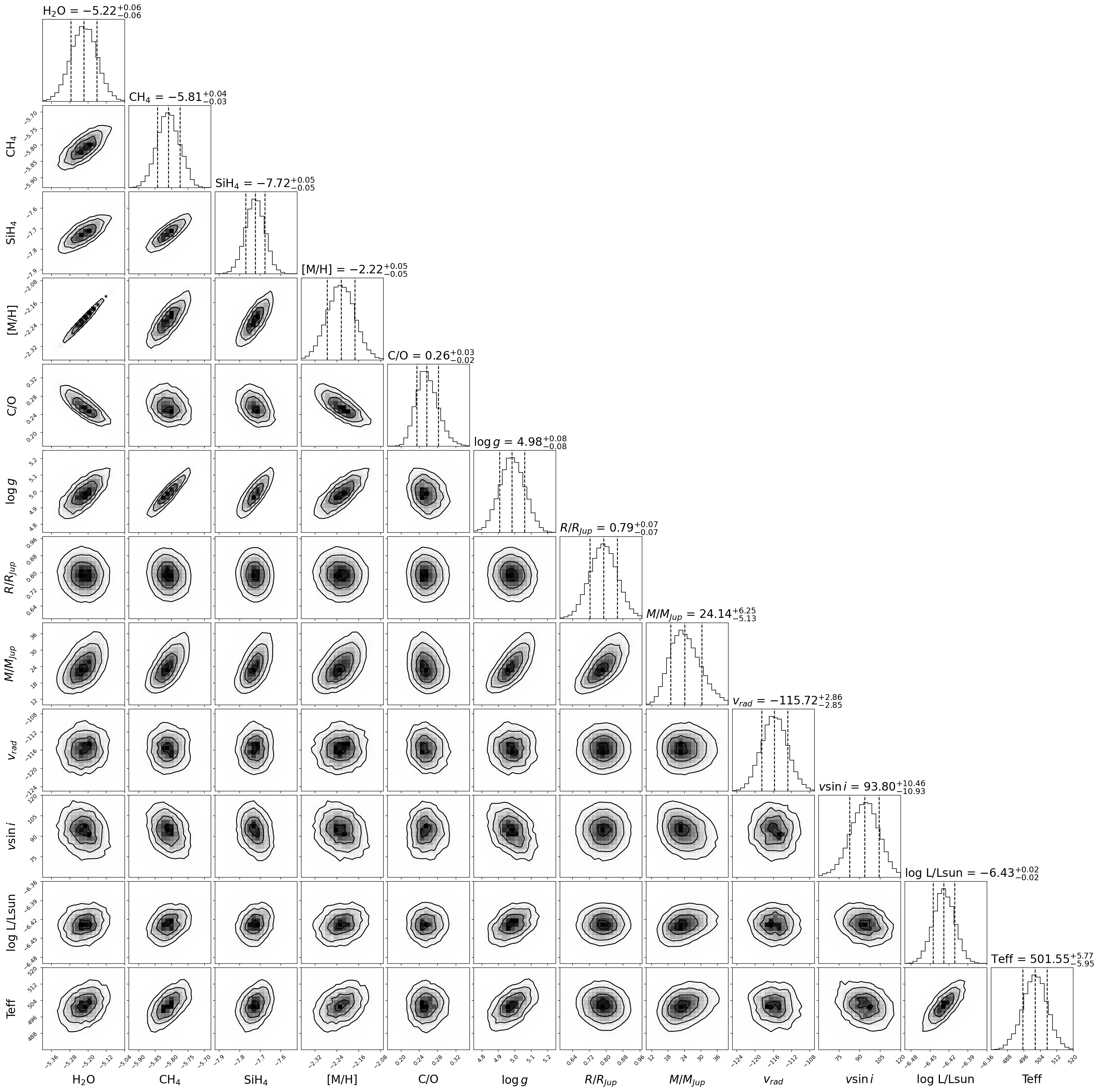} 
\caption{{\bf Retrieved and derived parameters for W1534.}
Diagonal panels display the distributions of parameters marginalized over all other quantities, with dashed lines indicating the median and $\pm$1$\sigma$ range, labeled above each panel. Interior contour plots display parameter correlations.
Abundances, surface gravity ({\logg}), radial velocity ($v_{rad}$), and rotational velocity ($v\sin{i}$) 
are retrieved from the posterior distributions; metallicity ([M/H]), C/O ratio, radius (R), mass (M), luminosity (log(L/L$_{sun}$)),
and effective temperature ({\teff}) are computed from these quantities.}
\label{fig:w1534corner}
\end{figure}

\begin{figure}[t]
\centering
\includegraphics[width=.8\textwidth]{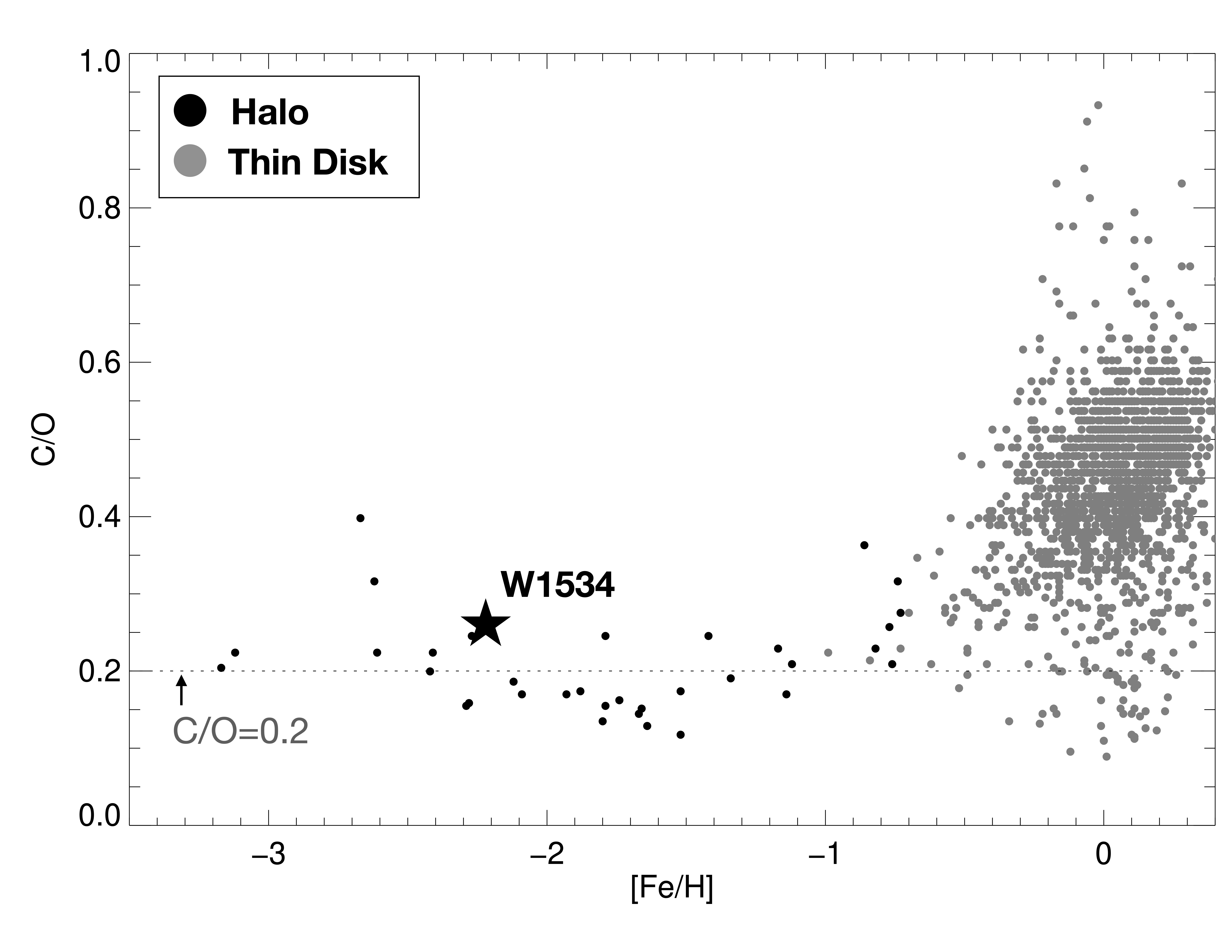} 
\caption{{\bf C/O ratio vs metallicity for FGK stars in the disk and halo.}
Plotted are the C/O values and metallicity [Fe/H] for FGK primarily disk stars as measured by \citep{Brewer16} and halo stars as measured by \citep{Akerman2004}. The retrieved values for W1534 are overplotted as a five point
star and labeled.}
\label{fig:CO}
\end{figure}

Combining the radial velocity with the proper motion and parallax of W1534 we determine Galactic velocity components relative to the Sun (not corrected to the Local Standard of Rest) of 
(U,V,W)=($-$78$\pm$4, $-$192$\pm$15, $-$116$\pm$4) km s$^{-1}$, for
a total space velocity of 238$\pm$14 km s$^{-1}$. Measurement errors were propagated with a 10$^{5}$-elements Monte Carlo. Synthetic values for the proper motion, radial velocity and parallax were drawn with four independent normal distributions following the measurement and uncertainty, and UVW values were calculated for each Monte Carlo element. The median and median absolute deviations were taken as the final measurement and error for the resulting 10$^{5}$ values for each space velocity direction.
Extended Data Figure~\ref{fig:Toomre} shows a Toomre diagram of U,V,W velocities for stars drawn from the Gaia Catalog of Nearby Stars matched with corresponding Gaia DR3 radial velocities (\citep{GCNS21}, \citep{GaiaDR3}). Objects are color-coded by their Galactic population following \citep{Nissen2004} with 
thin disk objects having V$_{tot}$ $\leq$ 85 km s$^{-1}$, 
thick disk objects having 85 $<$ V$_{tot}$ $\leq$ 180 km s$^{-1}$, 
and halo objects having V$_{tot}$ $>$ 180 km s$^{-1}$. W1534 sits squarely within the locus of halo objects, implying an age of 10--12 Gyr (\citep{Jofre2011}).

We note that Extended Data Figure~\ref{fig:w1534corner} also shows that the retrieval provided a vsini value of $\sim$94$\pm$11 km s$^{-1}$.  We used the published JWST instrumental profile for NIRSpec within the retrieval to broaden the lines.  We tested changing that instrumental profile to capture its influence on the results.  That change led to negligible differences in all parameters except vsini which we found to be degenerate with the chosen instrumental profile. As W1534 is exceedingly old, we expect this brown dwarf to have a relatively large vsini value, though we encourage the reader to view that value with caution as vsini values have yet to be calibrated with JWST data.

\subsection{Spectral Energy Distribution Construction and Results} \label{sec:SED}
A distance-calibrated spectral energy distribution (SED) inclusive of the prior literature and newly-obtained JWST data was generated using the open-source package SEDkit (\cite{Filippazzo20}), developed specifically to evaluate brown dwarf fundamental parameters.  All photometry used in the SED is listed in Table~\ref{tab:target}.  
Following \cite{Filippazzo15}, we first combined and scaled the NIRSpec spectra (prism and G395H) and photometry (ground based, Spitzer, WISE, and JWST/MIRI) to absolute fluxes using the measured parallax (\cite{Kirkpatrick21b}).  For data shortward of that collected, SEDkit linearly interpolates to zero.  For data longward of that collected, SEDkit appends a Rayleigh-Jeans tail out to 1000$\mu$m. SEDkit was then used to integrate the SED and calculate the bolometric luminosity ($L_{\rm bol}$) -- see \cite{Filippazzo15} for all procedural details.  Figure~\ref{fig:spectra} shows the output SED for W1534 between 0.9 - 5.2 $\mu$m (MIRI photometric points not shown). The measured bolometric luminosity for this source was log($L_{\rm bol}$/$L_{\odot}$)=	$-$6.3$\pm$0.1 which is consistent with the retrieval result to just over 1$\sigma$.

\subsection{Retrieval Analysis}
\label{sec:brewster}
We carried out a retrieval analysis of the NIRSpec G395H data using the publicly-available {\it Brewster} package (\cite{Burningham17,burningham2021}), developed to model substellar atmospheres. {\it Brewster} has been applied to a range of brown dwarf atmospheres, from cool Y dwarfs to warm L type subdwarfs, including cloudy and planetary-mass objects \citep{Burningham17,Gonzales18,Gonzales20,burningham2021,Calamari22,Gaarn2022,Vos22, Faherty24}. 

\subsubsection{Retrieval method}

{\it Brewster} consists of a forward model coupled to a Bayesian posterior sampling algorithm, in this case {\sc emcee} \citep{emcee}. Details on how {\it Brewster} operates (e.g. constraints on the thermal profile, the radiative transfer process, model convergence strategies, etc.) can be found in \citep{Faherty24}.  Here we detail only those adaptations that were made for the retrieval conducted in this study. 
  
The fit parameters and their associated priors are listed in Extended Data Table~\ref{tab:priors}. As with previous studies we've used 16 walkers per parameter and in this case with 29 parameters that means we had 464 walkers in our cloud-free retrieval. The parameters for the forward model are organized into gas-phase opacities, cloud opacity, temperature structure, and global properties of the target.

The gas-phase opacities are set by the choice of absorbing gases, their concentration, and vertical distribution in the model atmosphere. The opacities were derived from the compendium of \citep{freedman2008}, \citep{freedman2014}, appended in this work with SiH$_{4}$ from \citep{Grimm2021}, 12CO and 13CO opacities from \citep{Li2015} and \citep{Rothman2010}, and otherwise following the formulation with references there-in of \citep{Faherty24}. We included the following gases in this study: \hho, \chhhh, CO, \coo, \nhhh, \hhs, CH$_{3}$D, $^{12}$CO, $^{13}$CO, SiH$_{4}$, and \phhh. Concentrations (volume mixing ratios)
of the absorbing gases were assumed to be vertically constant as in previous studies with this code, which provides flexibility in arriving at possible solutions while retaining computational simplicity. Chemical equilibrium predictions for the gases (e.g \citep{Visscher2010}, \citep{Visscher12}) explored here support this approach. As with some previous applications of {\it Brewster} on cold brown dwarfs, we neglected cloud opacity. The fact that we are seeing to such deep levels of the atmosphere (see contribution function on the right panel of Figure~\ref{fig:retrievedprofile}) is suggestive that clouds are not present over the observable photosphere.  We leave the inclusion of clouds in cold brown dwarf atmospheres - either water ice, sulfide salt, or P-bearing clouds -- for a future study. 
 
\subsubsection{Retrieval results: Silane Detection}
We ran two models for W1534: one with SiH$_{4}$ gas opacity included and one without.  To distinguish the preferred model, we calculated the Bayesian Information Criterion (BIC) for both cases and compared. For a set of models, the one with the smallest (typically most negative) BIC will be the preferred or ``winning'' model,
while the strength of preference among lower-ranked models depends on the value of the difference in BICs relative to this best-fit model (\cite{kass1995}):

\begin{itemize}
  \item $0  < \Delta BIC < 2$: no significant preference;
  \item $2  < \Delta BIC < 6$: positive;
  \item $6  < \Delta BIC < 10$: strong;
 \item $10  < \Delta BIC$: very strong.
\end{itemize}

For our two models of W1534, the model that includes SiH$_{4}$ is strongly preferred over the one without it: BIC with SiH$_{4}$ = -84273, BIC without SiH$_{4}$ = -83383, $\Delta$BIC = 890.

\subsubsection{Retrieval results: thermal profile}
The retrieved thermal profile in Figure~\ref{fig:retrievedprofile} is best constrained in the visible photosphere at pressures between 1--100 bar.  Compared to retrievals for other cold brown dwarfs (e.g. \cite{Faherty24,Lew2024,Kothari2024}), the photosphere of W1534 probes nearly 10 times deeper.  The only comparable depths probed are for (1) the L subdwarf SDSS J125637.13-022452.4 (\cite{Gonzales2021}) -- though that was done only on the 1 - 2.5 $\mu$m coverage which probes warmer temperatures at deeper parts of the atmosphere --, and (2) the T subdwarf Wolf~1130C (Burgasser et al. submitted) -- which is warmer than W1534 but also low metallicity $[M/H]$ $\sim$ 0.7.
Examining the contribution function in the right panel of Figure~\ref{fig:retrievedprofile}, we see the methane spans the largest pressure range with the silane found at a deeper pressure.  As noted in the main text, the thermal profile of W1534 matches predictions from the 500K and 525K LowZ (\citep{Meisner2021}) self-consistent models through a portion of the visible photosphere. Briefly, the LowZ grid are a subset of non-irradiated models produced by the ScCHIMERA 1D-RCTE (radiative-convective thermochemical equilibrium) code.  The original ScCHIMERA tool was described and validated in \citep{Piskorz2018} and has been used in numerous subsequent exoplanet works (e.g., \citep{Arcangeli2018}; \citep{Gharib-Nezhad2019}; \citep{Welbanks2024}). At $\sim$ 10 bar, the retrieved profile has a ``kink" where the shape changes.   Below $\sim$ 10 bar the profile deviates significantly from an adiabat.  Given that the condensation curves of NH$_{4}$H$_{2}$PO$_{4}$ and {\hho} cross the thermal profile just above 1 bar, and ZnS and KCl cross just below 10 bar, a cloud may be a possible cause of the profile's deviation.   We ran a simple retrieval test using a grey cloud that forced an adiabatic profile for the deep atmosphere and found that it was also able to fit the data with no impact on retrieved abundances.  All abundances from the cloudy model were within 1$\sigma$ of the cloud-free model and the retrieved pressure for the cloud $\tau$=1 layer was placed at 29$_{-4}^{+3}$ bar. Detailed modeling of the potential cloud is outside the scope of this work but will be the subject of a future study.

\subsubsection{Retrieval results: fundamental parameters and abundances}
In order to calculate log($L_{\rm bol}$/$L_{\odot}$) with {\it Brewster}, we use the best fit distance calibrated model, extrapolate from $\sim$0.5 - 20 $\mu$m and then integrate under the curve.  Our retrieval result for log($L_{\rm bol}$/$L_{\odot}$) matches to within just over 1$\sigma$ with the SED analysis (see Extended Data Table~\ref{tab:target}).  The retrieved fundamental parameters of mass and $T_{\rm eff}$ also match prior estimates for this source as a very cold and metal-poor brown dwarf (e.g. \citep{Kirkpatrick21}). The retrieved {\teff} is comparable to that of solar-metallicity brown dwarfs around the T dwarf/Y dwarf transition (see \cite{Beiler24}).
More remarkable is the extreme subsolar metallicity inferred from the retrieved abundances, [M/H] = $-$2.22$\pm$0.05, which is comparable to the inferred metallicity of the early L-type ultrasubdwarf SDSS~J010448.46+153501.8 (\cite{Zhang2017}), which is likely a low-mass metal-poor star. Combined with its clear halo kinematics,
W1534 is the lowest-metallicity and likely oldest brown dwarf discovered to date.  

In Extended data table~\ref{tab:Abundances} we provide the retrieved gas abundances of H$_{2}$O, CH$_{4}$, and SiH$_{4}$ as well as the upper limits for undetected gases.  As a sanity check on the plausibility of the upper limits, we scaled the \citep{Lodders2002} abundances to the [M/H] and C/O ratio along the atmospheric profile for W1534. The resultant grid predictions are consistent with our 3$\sigma$ upper limits for H$_{2}$S, PH$_{3}$, CO and CO$_{2}$ but fall just outside for NH$_{3}$ (although the upper limit is consistent with our 4$\sigma$ value). This could reflect a peculiar nitrogen abundance or some hitherto unmodelled effect of low-metallicity chemistry.

\subsection{Silane Chemistry}
\begin{figure}[t]
\centering
\includegraphics[width=1.0\textwidth]{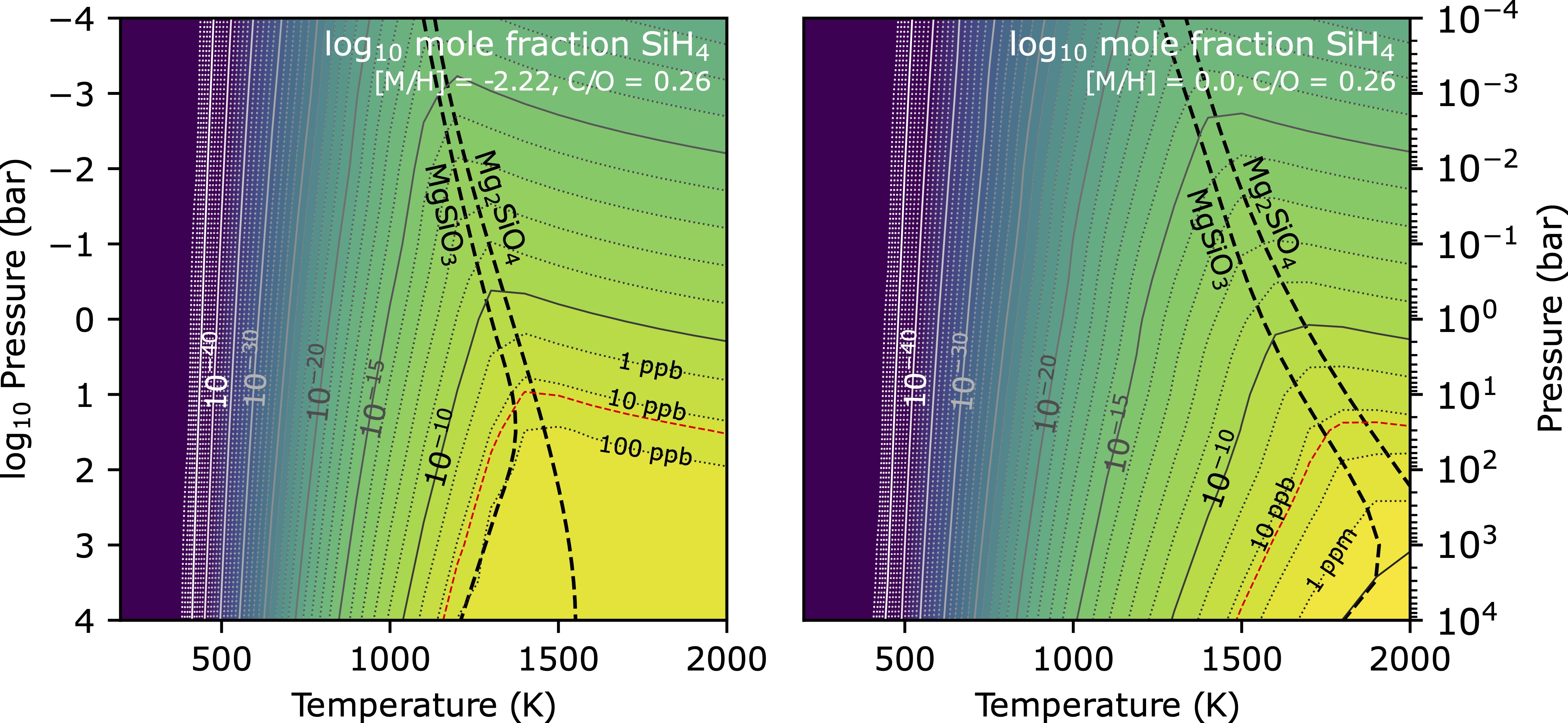} 
\caption{{\bf The equilibrium abundance of SiH$_4$ as a function of pressure and temperature with C/O=0.26 in low-metallicity (left) and solar-metallicity (right) gas.} In each panel the contours represent the mole fraction abundance of silane; the red dashed curve indicates the $\sim$20 ppb obtained from the atmospheric retrieval of W1534. The condensation curves for forsterite and enstatite are overplotted as dashed lines. At thermochemical equilibrium, SiH$_4$ is the most abundant Si-bearing gas at high temperatures and high pressures until it is removed at lower temperatures by silicate condensation, or replaced at lower pressures by other Si-bearing gases such as SiO (cf. \cite{Visscher2010}). }
\label{fig:metallicity-comparison}
\end{figure}

Thermochemical equilibrium models predict SiH$_4$ to be among the more abundant Si-bearing gases in the deep atmospheres of cool substellar objects, but like other Si species it is removed from the gas phase by the condensation of Mg silicates at depth (\cite{Fegley94,Visscher2010}). However, as shown in Extended Data Figure \ref{fig:metallicity-comparison} a reduced metallicity increases the range of pressures and temperatures over which SiH$_4$ is the dominant Si-bearing gas,  and significantly lowers the temperature at which Si-bearing gases are removed by silicate condensation. Furthermore, sufficiently rapid atmospheric mixing may quench SiH$_4$ oxidation to yield detectable photospheric SiH$_4$ abundances. 

Here we explored the possibility of SiH$_4$ quenching in the atmosphere of W1534 by using a timescale approximation for SiH$_4$$\rightarrow$SiO conversion kinetics. Following the general approach of \citep{Fegley94}, the oxidation of SiH$_4$ to SiO (and thus to subsequent oxides or silicates) plausibly proceeds via the series of reactions:
\begin{align}
    \rm H_2 & \rightarrow \rm 2H\\
        \rm H + SiH_4 & \rightarrow \rm SiH_3 + H_2\\
        \rm H + SiH_3 & \rightarrow \rm SiH_2 + H_2\\
        \rm H_2O + SiH_2 & \rightarrow \rm H_2Si{=}O + H_2 \label{rxn: H2SiO}\\
        \rm H_2Si{=}O + H_2O & \rightarrow \rm HSiOOH + H_2\\
        \rm HSiOOH & \rightarrow \rm SiO + H_2O\\ 
        \rm SiH_4 + H_2O & \rightarrow \rm SiO + 3H_2 \tag{Net}
\end{align}
The quenching behavior of the SiH$_4$-SiO system is analogous to quenching of the CO-CH$_4$ system, in that reactions among reduced species (SiH$_4$, SiH$_3$, SiH$_2$, etc.) and oxidized species (SiO, SiO$_2$, etc.) are expected to proceed much more rapidly than reactions between these groups (e.g., the formation of Si-O bonds). Noting the formation of the Si=O double bond, \cite{Fegley94} identify reaction (\ref{rxn: H2SiO}) as the rate-limiting step and adopt a rate constant from Figure 2 of \cite{Zachariah93}. A more recent revision to the rate constant for reaction (\ref{rxn: H2SiO}) is provided in Figure 3 of \cite{Zachariah95} and is approximated by the expression
\begin{equation}
    k_{\ref{rxn: H2SiO}}\approx 1.01\times10^{-12}\exp{(-5187.5/T)}{\rm~cm^{3}~s^{-1}}.
\end{equation}
The chemical timescale for SiH$_4$ is thus given by
\begin{equation}
    t_{chem}=\frac{[{\rm SiH_4}]}{k_{\ref{rxn: H2SiO}}[\rm SiH_2][H_2O]}.
\end{equation}
The quench condition for SiH$_4$ oxidation is approximated by
\begin{equation}
    t_{chem}{\rm(SiH_4)} \geq t_{mix} \approx \frac{H^2}{K_{zz}},
\end{equation}
where $H$ is the pressure scale height and $K_{zz}$ is the eddy diffusion coefficient. The quench level is defined where $t_{chem} = t_{mix}$. Above the quench level (i.e., at lower temperatures and pressures), SiH$_4$$\rightarrow$SiO reaction kinetics are too slow to keep up with atmospheric mixing ($t_{chem} > t_{mix}$) and the abundance of SiH$_4$ present at the quench level is mixed upward into the observable atmosphere.

\begin{figure}[t]
\centering
\includegraphics[width=.8\textwidth]{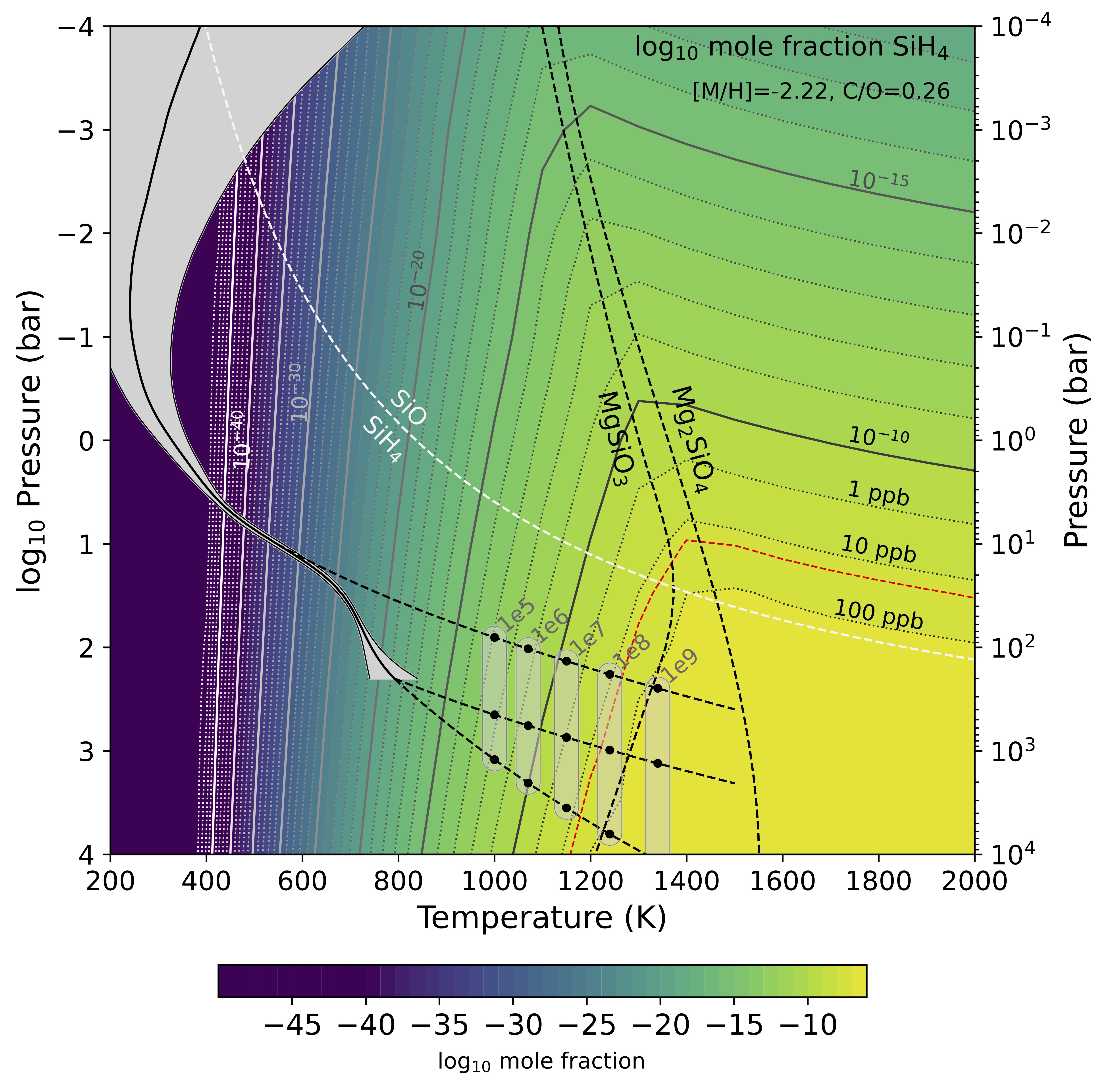} 
\caption{{\bf The retrieved thermal profile for W1534 extended to deeper pressures to extrapolate to the SiH$_4$ quench region.} The black line with grey error bar represents the retrieved thermal profile for W1534.  The colored contours represent the equilibrium abundance of silane calculated for an atmosphere with [M/H]=-2.22 and C/O=0.26, with the red dashed line corresponding to the $\sim$20 ppb obtained from the retrieval.  The condensation curves of enstatite and forsterite are overplotted, and demonstrate the rapid decrease in the SiH$_4$(g) abundance at lower temperatures (i.e., above the cloud base).  The thermal profile was extended to deeper pressures in three ways: by extrapolation from the trend of the retrieved profile (lower dashed line), an adiabatic extension from the deepest point of the retrieved profile (middle dashed line), and an adiabatic extension from the $\sim$ 10 bar level of the retrieved profile (top dashed line). The white dashed curve indicates the SiO-SiH$_4$ equal abundance boundary (cf. \cite{Visscher2010}). Highlighted as filled ovals are the vertical mixing rates parameterized by $K_{zz}$ (labelled in units of cm$^2$ s$^{-1}$) that indicate where quenching will occur along each thermal profile. } 
\label{fig:silane}
\end{figure}

Adopting this quench approximation for SiH$_{4}$, we examined the retrieved temperature-pressure profile for W1534 to determine the atmospheric mixing rates that would be required to obtain the retrieved abundance in the visible photosphere.  Extended Data Figure~\ref{fig:silane} shows a colored contour plot of silane abundance overplotted with the retrieved thermal profile from Figure~\ref{fig:retrievedprofile}.  Also shown are the condensation curves for the primary Si-bearing condensates MgSiO$_{3}$ and Mg$_{2}$SiO$_{4}$ (dashed, labeled lines).  The red contour line represents the retrieved abundance of 20 ppb SiH$_4$ in this work.  In order to reach the expected SiH$_4$ quench region we extrapolated the pressure-temperature profile from Figure~\ref{fig:retrievedprofile} to higher pressures and temperatures in three ways: (1) extending directly from the trend of the retrieved thermal profile (lower dashed line), 
(2) extending as an adiabat from the end of the retrieved thermal profile, (middle dashed line), 
or (3) extending as an adiabat from the $\sim$10 bar pressure where we see the profile ``kink'' and deviation from self-consistent model profiles (top dashed line). All three of these extensions cross the retrieved silane abundance above (i.e., at lower temperatures than) the enstatite and forsterite cloud condensation curves. 

\begin{figure}[t]
\centering
\includegraphics[width=1.0\textwidth]{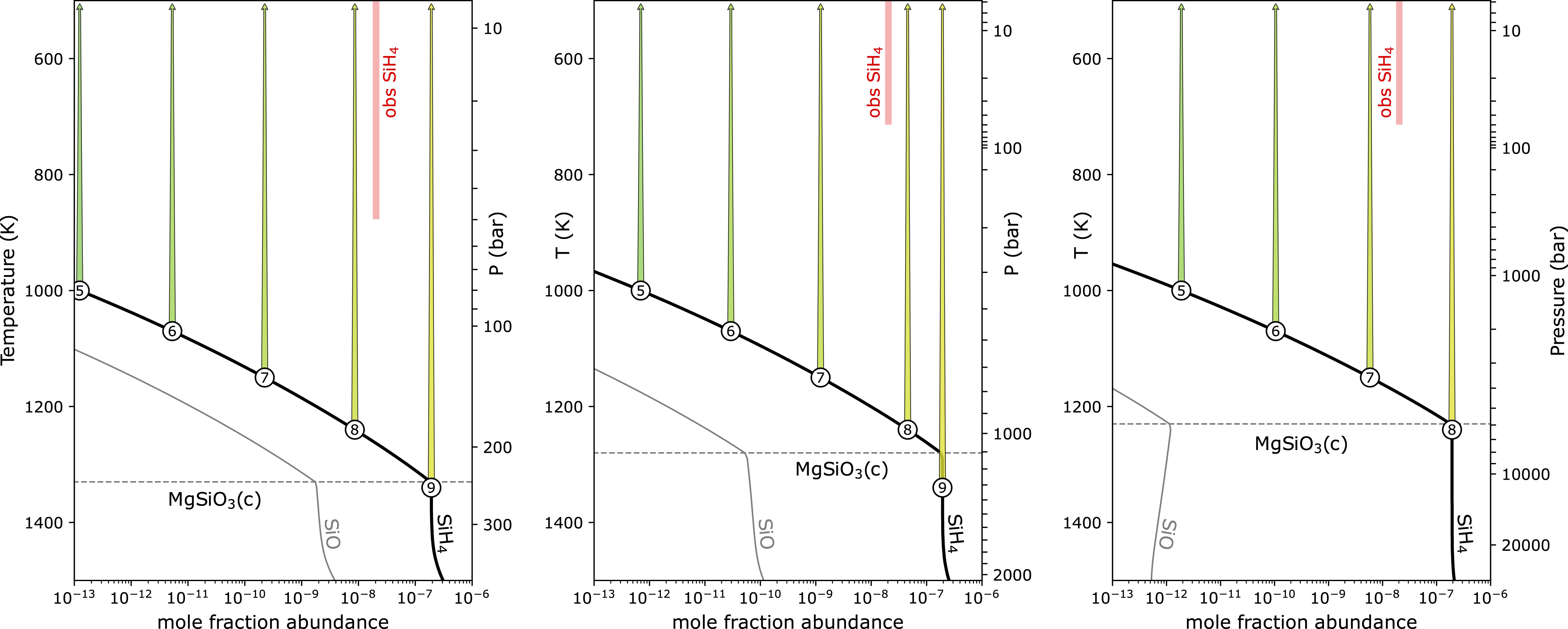} 
\caption{{\bf Quenching of silane along the thermal profile of W1534.} Equilibrium abundances of SiH$_4$ and SiO are shown along each of the extended thermal profiles in Figure \ref{fig:silane}, from the lowest pressure profile (left) to the highest pressure profile (right). Note the different pressure scale for each panel (right axes). The red band indicates the abundance and approximate observed altitude of the retrieved SiH$_4$. Under equilibrium conditions, the abundances of Si-bearing species rapidly decrease above the silicate cloud layer (horizontal dashed line).  However, transport-induced quenching may deliver much higher SiH$_4$ abundances into the upper atmosphere, depending upon the strength of atmospheric mixing. The circles along the SiH$_4$ profile label the values of log$_{10}$ $K_{zz}$ (cm$^2$ s$^{-1}$) that lead to the quenched abundances extending vertically upward to lower pressures.} 
\label{fig:silane_2d_mix}

\end{figure}

Extended Data Figure \ref{fig:silane} shows that the depth from which SiH$_4$ quenching occurs depends on the atmospheric mixing rate, parameterized by $K_{zz}$. A corresponding Extended Data Figure \ref{fig:silane_2d_mix} shows the SiH$_4$ abundance and its quench points \textit{along} each of the atmospheric profiles for a range of $K_{zz}$ values. These results indicate that the retrieved SiH$_4$ abundance can be produced by transport-induced quenching for $K_{zz}$ values greater than $\sim10^{7}-10^{8}$ cm$^{2}$ s$^{-1}$, depending upon the adopted thermal profile. Moreover, in each case the model results suggest that SiH$_4$ is quenched from just above the MgSiO$_3$ cloud layer, but before it has been completely depleted by oxidation and removal into the silicate clouds.

\newpage
\backmatter
\bmhead{Availability of data and materials} The JWST data in this paper are part of GO program 3558 (PI Meisner) and will become publicly available in the Barbara A. Mikulski Archive for Space Telescopes (MAST; \url{https://archive.stsci.edu}) under that program ID in July 2025. 

\bmhead{Code availability} The data reduction pipeline jwst can be found at \url{https://jwst-pipeline.readthedocs.io/en/latest/. }
This study made use of version 1.15.1 of the pipeline which is available at \url{https://zenodo.org/records/12692459}
The {\it Brewster} code is open-source and available at the following GitHub repository: \url{https://github.com/substellar/brewster}.  
The SEDkit code is open-source and available at \url{https://github.com/hover2pi/sedkit}.

\bmhead{Acknowledgments} 
JF acknowledges funding from the Heising Simons Foundation as well as NSF award \#2238468, \#1909776, and NASA Award \#80NSSC22K0142, and support from JWST-GO-03558. BB acknowledges support from UK Research and Innovation Science and Technology Facilities Council [ST/X001091/1].  CV acknowledges support from JWST-AR-2232. AJB acknowledges funding support from the Heising Simons Foundation. ELM acknowledges funding support from the European Research Council Advanced grant Substellar project number 101054354.  BG acknowledges support from the Polish National Science Center (NCN) under SONATA grant No. 2021/43/D/ST9/0194. MRL acknowledges support from JWST-GO-03558. VJSB, NL, EM and MRZO acknowledges support from grant PID2022-137241NB-C4[1,2] funded by Agencia Estatal de Investigación of the  Ministerio de Ciencia, Innovación y Universidades (MICIU/AEI/10.13039/501100011033) and ERDF/EU. For the purpose of open access, the author has applied a Creative Commons Attribution (CC BY) licence to any Author Accepted Manuscript version arising. Portions of this research were carried out at the Jet Propulsion Laboratory, California In- stitute of Technology, under a contract with the National Aeronautics and Space Administration.

\bmhead{Author contributions}
AM was PI of the JWST proposal and observing execution.  JF oversaw all data reduction, analysis and modeling.  
BB completed the atmospheric retrieval.  JG extracted the radial velocity of W1534. GS, AR, and SA contributed to data reduction and SED analysis. CV completed the chemistry models and quenching kinetics analysis. MRL first identified the SiH$_4$ feature and suggested the initial idea/motivation for the manuscript. AJB, ELM, AS, DC, JDK, DB,  PE, EG, FM, SL, NL, SC, PT, MC, MRZO, VJSB, BG, EW,MP, and JZ  contributed to interpretation of the results and editing of the manuscript.

\bmhead{Author Information} The authors declare no competing interests.  Supplementary Information is available for this paper. Correspondence and requests for materials should be addressed to JF (jfaherty@amnh.org). 

\clearpage

\begin{table}[t]
\begin{center}
\begin{minipage}{1.0\textwidth}
\caption{Parameters of interest for W1534}\label{tab:target}
\begin{tabular*}{\textwidth}{@{\extracolsep{\fill}}llllll@{\extracolsep{\fill}}}
\toprule
    \bf{Astrometry} & \bf{W1534}  &\bf{units}   & \bf{ref}       \\
\midrule	
    $\alpha$ (J2000)         &    233.621873             	& deg             	& \citep{Kirkpatrick21b}\\
    $\delta$ (J2000)         &     $-$10.721775                        	& deg            	&  \citep{Kirkpatrick21b} \\
    $\varpi$           & 61$\pm$5               	 & mas                   &  \citep{Kirkpatrick21b} \\
    Distance         &    16.3$^{+1.4}_{-1.2}$           	& pc                	& \citep{Kirkpatrick21b}   \\
    $\mu_{\alpha}$    & $-$1253$\pm$9     & mas yr$^{-1}$     	& \citep{Kirkpatrick21b} \\
    $\mu_{\delta}$     & $-$2377$\pm$7     & mas yr$^{-1}$     	& \citep{Kirkpatrick21b}\\
   X   & 	13.3 $\pm$ 1.14		& pc		& This paper\\
   Y   & 	$-$1.22 $\pm$  0.11		& pc		& This paper \\
   Z   & 9.4 $\pm$ 0.8			& pc		& This paper\\
\toprule
    \bf{Kinematics}\\
\midrule	
   V$_{\rm tan}$ 	&	207$\pm$18		& km\,s$^{-1}$		& \citep{Kirkpatrick21b}\\
   RV 	&	$-$116$\pm$4		& km\,s$^{-1}$		& This study\\
   U   &  $-$78$\pm$4			& km\,s$^{-1}$		& This study\\
   V   & $-$192$\pm$15			& km\,s$^{-1}$		& This study\\
   W   &  $-$116$\pm$4			& km\,s$^{-1}$		& This study\\
   Total Velocity & 238$\pm$14 & km\,s$^{-1}$		& This study\\
      Age	&	10 - 12		& Gyr			& This study\\
\toprule
    \bf{Photometry}\\
\midrule	
MKO $Y$ & $>$ 21.79 & mag & \citep{Meisner23RN}\\
     MKO $J$				& 24.5$\pm$0.3		& mag			& \citep{Meisner23} \\
    MKO $H$				&$>$18.6			& mag			&\citep{Meisner20} \\
    $K_{s}$				&$>$17.9				& mag			& \citep{Meisner20}\\
    F110W				&24.70$\pm$0.08				& mag			& \citep{Kirkpatrick21b}\\
    WISE W1				& 18.18$\pm$0.19 			& mag			& \citep{Marocco21}\\
    WISE W2				& 16.15$\pm$0.08			& mag			& \citep{Marocco21}\\
    Spitzer [3.6]		& 16.69$\pm$0.03 			& mag			& \citep{Meisner20}\\
   Spitzer [4.5]		& 15.77$\pm$0.02			& mag			& \citep{Meisner20} \\
    JWST F1500W	 &14.12$\pm$0.06			& mag			& This study \\
    JWST F1800W		&	13.76$\pm$0.06		& mag			& This study \\
    JWST F2100W &	13.58$\pm$0.06		& mag			& This study \\
 \toprule
    \bf{Synthetic Photometry from NIRSpec prism}\\
\midrule	
MKO $Y$ & 25.33$\pm$0.07& mag			& This study \\
MKO $J$ & 24.34$\pm$0.13& mag			& This study \\
MKO $H$ & 22.53$\pm$0.05& mag			& This study \\
MKO $K$ & $> 25.7$ & mag			& This study \\
\toprule
    \bf{SED Fundamental Parameter}\\
\midrule	
   log($L_{\rm bol}$/$L_{\odot}$)&	$-$6.3$\pm$0.1 		& dex & This study \\
    \toprule
   \bf{Retrieved Fundamental Parameters}   \\        
\midrule
log($L_{\rm bol}$/$L_{\odot}$) & $-$6.43$\pm$0.02 & dex& This study\\
Radius & 0.79$_{-0.06}^{+0.07}$ &$R_{\rm Jup}$ & This study\\
Mass & $24^{+6}_{-5}$&$M_{\rm Jup}$& This study\\
$\log g$ & 4.98 $\pm$ 0.08& log cm~s$^{-2}$ & This study\\
 $T_{\rm eff}$ &	502$\pm6$ 		&K& This study\\
\hline
\end{tabular*}
\end{minipage}
\end{center}
\end{table}

\begin{table}
\caption{Parameters and priors adopted for the retrieval analysis. \label{tab:priors} }
\begin{tabular}{ll}
\toprule
Parameter & Prior \\
\midrule
gas fraction, ($X_{gas}$) & log-uniform, $\log X_{gas} \geq -12.0$, $\sum_{gas}{X_{gas}} \leq 1.0$ \\
thermal profile, $T$ & uniform, constrained by $0.0~{\rm K} < T_{i} < 5000.0~{\rm K}$ \\
profile smoothing parameter, $\gamma$ & uniform, $0 < \gamma < 10^5$ \\
gravity, $\log g$ & uniform, constrained by $1 \Mjup\  \leq gR^{2} / G \leq 80 \Mjup $ \\ 
scale factor, $R^{2} / D^{2}$ & uniform, constrained by $0.5 \Rjup\ \leq R \leq 2.0 \Rjup$ \\
rotational velocity, $v \sin i$ & uniform, $0\ {\rm km s^{-1}} < v\sin i < 150\ {\rm km s^{-1}}$ \\
radial velocity, RV & uniform, $-250\ {\rm km s^{-1}} < RV < 250\ {\rm km s^{-1}}$  \\
tolerance factor, $b$ & uniform, $\log (0.01 \times min(\sigma_{i}^{2})) \leq b \leq \log(100 \times max(\sigma_{i}^{2}))$ \\
\botrule
\end{tabular}
\end{table}

\begin{table}[t]
\begin{center}
\begin{minipage}{\textwidth}
\caption{Retrieved Gas Abundances}\label{tab:Abundances}
\begin{tabular*}{\textwidth}{@{\extracolsep{\fill}}lccccc@{\extracolsep{\fill}}}
\toprule
\bf{Detected Gases}& 
\bf{W1534}&
\bf{units}\\
\midrule
$\log f_{gas}$ H$_{2}$O & $-5.22\pm0.06$ & dex \\
$\log f_{gas}$ CH$_{4}$ & $-$5.81$^{+0.04}_{-0.03}$ & dex  \\
$\log f_{gas}$ SiH$_{4}$ & $-7.72\pm0.05$ & dex \\
\hline
\toprule
\bf{Upper Limits for Undetected Gases\footnote{Upper limits taken as 97.6 (2$\sigma$ upper bound) percentiles in posterior}}\\
\midrule
$\log f_{gas}$ CH$_{3}$D & $\ltsimeq -4.87$& dex \\
$\log f_{gas}$ 12CO & $\ltsimeq -8.22 $& dex       \\
$\log f_{gas}$ 13CO & $\ltsimeq -8.52 $& dex       \\
$\log f_{gas}$ CO$_{2}$ &$\ltsimeq -10.06 $& dex  \\
$\log f_{gas}$ NH$_{3}$ &$\ltsimeq -6.95 $ & dex  \\
$\log f_{gas}$ H$_{2}$S &$\ltsimeq -5.60 $ & dex \\
$\log f_{gas}$ PH$_{3}$ & $\ltsimeq -8.88$ & dex \\

\hline
\toprule
\bf{Metallicity and C/O}\\
\midrule
$[M/H]$ & $-$2.22 $\pm{0.05}$ & dex \\
C/O & $0.26^{+0.03}_{-0.02}$ \\
\botrule
\end{tabular*}
\end{minipage}
\end{center}
\end{table}

\begin{thebibliography}{10}
\expandafter\ifx\csname url\endcsname\relax
  \def\url#1{\burl{#1}}\fi
\expandafter\ifx\csname urlprefix\endcsname\relax\def\urlprefix{URL }\fi
\providecommand{\bibinfo}[2]{#2}
\providecommand{\eprint}[2][]{\url{#2}}
\providecommand{\doi}[1]{\url{https://doi.org/#1}}
\bibcommenthead

\bibitem{Kirkpatrick24}
\bibinfo{author}{{Kirkpatrick}, J.~D.} \emph{et~al.}
\newblock \bibinfo{title}{{The Initial Mass Function Based on the Full-sky 20
  pc Census of {\ensuremath{\sim}}3600 Stars and Brown Dwarfs}}.
\newblock \emph{\bibinfo{journal}{\apjs}} \textbf{\bibinfo{volume}{271}},
  \bibinfo{pages}{55} (\bibinfo{year}{2024}).

\bibitem{Fegley94}
\bibinfo{author}{{Fegley}, J., Bruce} \& \bibinfo{author}{{Lodders}, K.}
\newblock \bibinfo{title}{{Chemical Models of the Deep Atmospheres of Jupiter
  and Saturn}}.
\newblock \emph{\bibinfo{journal}{\icarus}} \textbf{\bibinfo{volume}{110}},
  \bibinfo{pages}{117--154} (\bibinfo{year}{1994}).

\bibitem{Fletcher2016}
\bibinfo{author}{{Fletcher}, L.~N.} \emph{et~al.}
\newblock \bibinfo{title}{{Mid-infrared mapping of Jupiter's temperatures,
  aerosol opacity and chemical distributions with IRTF/TEXES}}.
\newblock \emph{\bibinfo{journal}{\icarus}} \textbf{\bibinfo{volume}{278}},
  \bibinfo{pages}{128--161} (\bibinfo{year}{2016}).

\bibitem{Atreya2005}
\bibinfo{author}{{Atreya}, S.~K.}, \bibinfo{author}{{Wong}, A.~S.},
  \bibinfo{author}{{Baines}, K.~H.}, \bibinfo{author}{{Wong}, M.~H.} \&
  \bibinfo{author}{{Owen}, T.~C.}
\newblock \bibinfo{title}{{Jupiter's ammonia clouds{\textemdash}localized or
  ubiquitous?}}
\newblock \emph{\bibinfo{journal}{\planss}} \textbf{\bibinfo{volume}{53}},
  \bibinfo{pages}{498--507} (\bibinfo{year}{2005}).

\bibitem{Wong04}
\bibinfo{author}{{Wong}, M.~H.}, \bibinfo{author}{{Mahaffy}, P.~R.},
  \bibinfo{author}{{Atreya}, S.~K.}, \bibinfo{author}{{Niemann}, H.~B.} \&
  \bibinfo{author}{{Owen}, T.~C.}
\newblock \bibinfo{title}{{Updated Galileo probe mass spectrometer measurements
  of carbon, oxygen, nitrogen, and sulfur on Jupiter}}.
\newblock \emph{\bibinfo{journal}{\icarus}} \textbf{\bibinfo{volume}{171}},
  \bibinfo{pages}{153--170} (\bibinfo{year}{2004}).

\bibitem{Gapp24}
\bibinfo{author}{{Gapp}, C.} \emph{et~al.}
\newblock \bibinfo{title}{{Abundances of trace constituents in Jupiter's
  atmosphere inferred from Herschel/PACS observations}}.
\newblock \emph{\bibinfo{journal}{\aap}} \textbf{\bibinfo{volume}{688}},
  \bibinfo{pages}{A10} (\bibinfo{year}{2024}).

\bibitem{Luhman14}
\bibinfo{author}{{Luhman}, K.~L.}
\newblock \bibinfo{title}{{Discovery of a \raisebox{-0.5ex}\textasciitilde250 K
  Brown Dwarf at 2 pc from the Sun}}.
\newblock \emph{\bibinfo{journal}{\apjl}} \textbf{\bibinfo{volume}{786}},
  \bibinfo{pages}{L18} (\bibinfo{year}{2014}).

\bibitem{Luhman24}
\bibinfo{author}{{Luhman}, K.~L.} \emph{et~al.}
\newblock \bibinfo{title}{{JWST/NIRSpec Observations of the Coldest Known Brown
  Dwarf}}.
\newblock \emph{\bibinfo{journal}{\aj}} \textbf{\bibinfo{volume}{167}},
  \bibinfo{pages}{5} (\bibinfo{year}{2024}).

\bibitem{Rowland24}
\bibinfo{author}{{Rowland}, M.}
\newblock \bibinfo{title}{{Protosolar D-to-H abundance and one part per billion
  PH$_{3}$ in the coldest brown dwarf}}.
\newblock \emph{\bibinfo{journal}{ApJL submitted}}  (\bibinfo{year}{2024}).

\bibitem{Matthews2024}
\bibinfo{author}{{Matthews}, E.~C.} \emph{et~al.}
\newblock \bibinfo{title}{{A temperate super-Jupiter imaged with JWST in the
  mid-infrared}}.
\newblock \emph{\bibinfo{journal}{\nat}} \textbf{\bibinfo{volume}{633}},
  \bibinfo{pages}{789--792} (\bibinfo{year}{2024}).

\bibitem{Lacy23}
\bibinfo{author}{{Lacy}, B.} \& \bibinfo{author}{{Burrows}, A.}
\newblock \bibinfo{title}{{Self-consistent Models of Y Dwarf Atmospheres with
  Water Clouds and Disequilibrium Chemistry}}.
\newblock \emph{\bibinfo{journal}{\apj}} \textbf{\bibinfo{volume}{950}},
  \bibinfo{pages}{8} (\bibinfo{year}{2023}).

\bibitem{Burgasser2007}
\bibinfo{author}{{Burgasser}, A.~J.}, \bibinfo{author}{{Cruz}, K.~L.} \&
  \bibinfo{author}{{Kirkpatrick}, J.~D.}
\newblock \bibinfo{title}{{Optical Spectroscopy of 2MASS Color-selected
  Ultracool Subdwarfs}}.
\newblock \emph{\bibinfo{journal}{\apj}} \textbf{\bibinfo{volume}{657}},
  \bibinfo{pages}{494--510} (\bibinfo{year}{2007}).

\bibitem{Gonzales18}
\bibinfo{author}{{Gonzales}, E.~C.}, \bibinfo{author}{{Faherty}, J.~K.},
  \bibinfo{author}{{Gagn{\'e}}, J.}, \bibinfo{author}{{Artigau}, {\'E}.} \&
  \bibinfo{author}{{Bardalez Gagliuffi}, D.}
\newblock \bibinfo{title}{{Understanding Fundamental Properties and Atmospheric
  Features of Subdwarfs via a Case Study of SDSS J125637.13-022452.4}}.
\newblock \emph{\bibinfo{journal}{\apj}} \textbf{\bibinfo{volume}{864}},
  \bibinfo{pages}{100} (\bibinfo{year}{2018}).

\bibitem{Visscher2010}
\bibinfo{author}{{Visscher}, C.}, \bibinfo{author}{{Lodders}, K.} \&
  \bibinfo{author}{{Fegley}, J., Bruce}.
\newblock \bibinfo{title}{{Atmospheric Chemistry in Giant Planets, Brown
  Dwarfs, and Low-mass Dwarf Stars. III. Iron, Magnesium, and Silicon}}.
\newblock \emph{\bibinfo{journal}{\apj}} \textbf{\bibinfo{volume}{716}},
  \bibinfo{pages}{1060--1075} (\bibinfo{year}{2010}).

\bibitem{Schneider2020}
\bibinfo{author}{{Schneider}, A.~C.} \emph{et~al.}
\newblock \bibinfo{title}{{WISEA J041451.67-585456.7 and WISEA
  J181006.18-101000.5: The First Extreme T-type Subdwarfs?}}
\newblock \emph{\bibinfo{journal}{\apj}} \textbf{\bibinfo{volume}{898}},
  \bibinfo{pages}{77} (\bibinfo{year}{2020}).

\bibitem{Lodieu22}
\bibinfo{author}{{Lodieu}, N.}, \bibinfo{author}{{Zapatero Osorio}, M.~R.},
  \bibinfo{author}{{Mart{\'\i}n}, E.~L.}, \bibinfo{author}{{Rebolo L{\'o}pez},
  R.} \& \bibinfo{author}{{Gauza}, B.}
\newblock \bibinfo{title}{{Physical properties and trigonometric distance of
  the peculiar dwarf WISE J181005.5{\ensuremath{-}}101002.3}}.
\newblock \emph{\bibinfo{journal}{\aap}} \textbf{\bibinfo{volume}{663}},
  \bibinfo{pages}{A84} (\bibinfo{year}{2022}).

\bibitem{Burgasser2024}
\bibinfo{author}{{Burgasser}, A.~J.} \emph{et~al.}
\newblock \bibinfo{title}{{New Cold Subdwarf Discoveries from Backyard Worlds
  and a Metallicity Classification System for T Subdwarfs}}.
\newblock \emph{\bibinfo{journal}{arXiv e-prints}}
  \bibinfo{pages}{arXiv:2411.01378} (\bibinfo{year}{2024}).

\bibitem{Linsky1969}
\bibinfo{author}{{Linsky}, J.~L.}
\newblock \bibinfo{title}{{On the Pressure-Induced Opacity of Molecular
  Hydrogen in Late-Type Stars}}.
\newblock \emph{\bibinfo{journal}{\apj}} \textbf{\bibinfo{volume}{156}},
  \bibinfo{pages}{989--+} (\bibinfo{year}{1969}).

\bibitem{Burgasser2003}
\bibinfo{author}{{Burgasser}, A.~J.} \emph{et~al.}
\newblock \bibinfo{title}{{The First Substellar Subdwarf? Discovery of a
  Metal-poor L Dwarf with Halo Kinematics}}.
\newblock \emph{\bibinfo{journal}{\apj}} \textbf{\bibinfo{volume}{592}},
  \bibinfo{pages}{1186--1192} (\bibinfo{year}{2003}).

\bibitem{Yamamura2010}
\bibinfo{author}{{Yamamura}, I.}, \bibinfo{author}{{Tsuji}, T.} \&
  \bibinfo{author}{{Tanab{\'e}}, T.}
\newblock \bibinfo{title}{{AKARI Observations of Brown Dwarfs. I. CO and
  CO$_{2}$ Bands in the Near-infrared Spectra}}.
\newblock \emph{\bibinfo{journal}{\apj}} \textbf{\bibinfo{volume}{722}},
  \bibinfo{pages}{682--698} (\bibinfo{year}{2010}).

\bibitem{Faherty24}
\bibinfo{author}{{Faherty}, J.~K.} \emph{et~al.}
\newblock \bibinfo{title}{{Methane emission from a cool brown dwarf}}.
\newblock \emph{\bibinfo{journal}{\nat}} \textbf{\bibinfo{volume}{628}},
  \bibinfo{pages}{511--514} (\bibinfo{year}{2024}).

\bibitem{Lew2024}
\bibinfo{author}{{Lew}, B. W.~P.} \emph{et~al.}
\newblock \bibinfo{title}{{High-precision Atmospheric Characterization of a Y
  Dwarf with JWST NIRSpec G395H Spectroscopy: Isotopologue, C/O Ratio,
  Metallicity, and the Abundances of Six Molecular Species}}.
\newblock \emph{\bibinfo{journal}{\aj}} \textbf{\bibinfo{volume}{167}},
  \bibinfo{pages}{237} (\bibinfo{year}{2024}).

\bibitem{Meisner20}
\bibinfo{author}{{Meisner}, A.~M.} \emph{et~al.}
\newblock \bibinfo{title}{{Expanding the Y Dwarf Census with Spitzer Follow-up
  of the Coldest CatWISE Solar Neighborhood Discoveries}}.
\newblock \emph{\bibinfo{journal}{\apj}} \textbf{\bibinfo{volume}{889}},
  \bibinfo{pages}{74} (\bibinfo{year}{2020}).

\bibitem{Kirkpatrick21b}
\bibinfo{author}{{Kirkpatrick}, J.~D.} \emph{et~al.}
\newblock \bibinfo{title}{{The Enigmatic Brown Dwarf WISEA J153429.75-104303.3
  (a.k.a. ``The Accident'')}}.
\newblock \emph{\bibinfo{journal}{\apjl}} \textbf{\bibinfo{volume}{915}},
  \bibinfo{pages}{L6} (\bibinfo{year}{2021}).

\bibitem{Beiler24}
\bibinfo{author}{{Beiler}, S.~A.} \emph{et~al.}
\newblock \bibinfo{title}{{Precise Bolometric Luminosities and Effective
  Temperatures of 23 late-T and Y dwarfs Obtained with JWST}}.
\newblock \emph{\bibinfo{journal}{arXiv e-prints}}
  \bibinfo{pages}{arXiv:2407.08518} (\bibinfo{year}{2024}).

\bibitem{GCNS21}
\bibinfo{author}{{Gaia Collaboration}} \emph{et~al.}
\newblock \bibinfo{title}{{Gaia Early Data Release 3. The Gaia Catalogue of
  Nearby Stars}}.
\newblock \emph{\bibinfo{journal}{\aap}} \textbf{\bibinfo{volume}{649}},
  \bibinfo{pages}{A6} (\bibinfo{year}{2021}).

\bibitem{GaiaDR3}
\bibinfo{author}{{Gaia Collaboration}} \emph{et~al.}
\newblock \bibinfo{title}{{Gaia Data Release 3. Summary of the content and
  survey properties}}.
\newblock \emph{\bibinfo{journal}{\aap}} \textbf{\bibinfo{volume}{674}},
  \bibinfo{pages}{A1} (\bibinfo{year}{2023}).

\bibitem{Bensby2014}
\bibinfo{author}{{Bensby}, T.}, \bibinfo{author}{{Feltzing}, S.} \&
  \bibinfo{author}{{Oey}, M.~S.}
\newblock \bibinfo{title}{{Exploring the Milky Way stellar disk. A detailed
  elemental abundance study of 714 F and G dwarf stars in the solar
  neighbourhood}}.
\newblock \emph{\bibinfo{journal}{\aap}} \textbf{\bibinfo{volume}{562}},
  \bibinfo{pages}{A71} (\bibinfo{year}{2014}).

\bibitem{Burningham17}
\bibinfo{author}{{Burningham}, B.} \emph{et~al.}
\newblock \bibinfo{title}{{Retrieval of atmospheric properties of cloudy L
  dwarfs}}.
\newblock \emph{\bibinfo{journal}{\mnras}} \textbf{\bibinfo{volume}{470}},
  \bibinfo{pages}{1177--1197} (\bibinfo{year}{2017}).

\bibitem{burningham2021}
\bibinfo{author}{{Burningham}, B.} \emph{et~al.}
\newblock \bibinfo{title}{{Cloud busting: enstatite and quartz clouds in the
  atmosphere of 2M2224-0158}}.
\newblock \emph{\bibinfo{journal}{\mnras}} \textbf{\bibinfo{volume}{506}},
  \bibinfo{pages}{1944--1961} (\bibinfo{year}{2021}).

\bibitem{Gonzales20}
\bibinfo{author}{{Gonzales}, E.~C.} \emph{et~al.}
\newblock \bibinfo{title}{{Retrieval of the d/sdL7+T7.5p Binary SDSS
  J1416+1348AB}}.
\newblock \emph{\bibinfo{journal}{\apj}} \textbf{\bibinfo{volume}{905}},
  \bibinfo{pages}{46} (\bibinfo{year}{2020}).

\bibitem{Vos22}
\bibinfo{author}{{Vos}, J.~M.} \emph{et~al.}
\newblock \bibinfo{title}{{Let the Great World Spin: Revealing the Stormy,
  Turbulent Nature of Young Giant Exoplanet Analogs with the Spitzer Space
  Telescope}}.
\newblock \emph{\bibinfo{journal}{\apj}} \textbf{\bibinfo{volume}{924}},
  \bibinfo{pages}{68} (\bibinfo{year}{2022}).

\bibitem{Calamari22}
\bibinfo{author}{{Calamari}, E.} \emph{et~al.}
\newblock \bibinfo{title}{{An Atmospheric Retrieval of the Brown Dwarf Gliese
  229B}}.
\newblock \emph{\bibinfo{journal}{\apj}} \textbf{\bibinfo{volume}{940}},
  \bibinfo{pages}{164} (\bibinfo{year}{2022}).

\bibitem{Meisner2021}
\bibinfo{author}{{Meisner}, A.~M.} \emph{et~al.}
\newblock \bibinfo{title}{{New Candidate Extreme T Subdwarfs from the Backyard
  Worlds: Planet 9 Citizen Science Project}}.
\newblock \emph{\bibinfo{journal}{\apj}} \textbf{\bibinfo{volume}{915}},
  \bibinfo{pages}{120} (\bibinfo{year}{2021}).

\bibitem{lodders2021}
\bibinfo{author}{{Lodders}, K.}
\newblock \bibinfo{title}{{Relative Atomic Solar System Abundances, Mass
  Fractions, and Atomic Masses of the Elements and Their Isotopes, Composition
  of the Solar Photosphere, and Compositions of the Major Chondritic Meteorite
  Groups}}.
\newblock \emph{\bibinfo{journal}{\ssr}} \textbf{\bibinfo{volume}{217}},
  \bibinfo{pages}{44} (\bibinfo{year}{2021}).

\bibitem{Treffers1978}
\bibinfo{author}{{Treffers}, R.~R.}, \bibinfo{author}{{Larson}, H.~P.},
  \bibinfo{author}{{Fink}, U.} \& \bibinfo{author}{{Gautier}, T.~N.}
\newblock \bibinfo{title}{{Upper limits to trace constituents in Jupiter's
  atmosphere from an analysis of its 5-{\ensuremath{\mu}}m spectrum}}.
\newblock \emph{\bibinfo{journal}{\icarus}} \textbf{\bibinfo{volume}{34}},
  \bibinfo{pages}{331--343} (\bibinfo{year}{1978}).

\bibitem{Noll1991}
\bibinfo{author}{{Noll}, K.~S.} \& \bibinfo{author}{{Larson}, H.~P.}
\newblock \bibinfo{title}{{The spectrum of Saturn from 1990 to 2230 cm $^{-1}$:
  Abundances of AsH $_{3}$, CH $_{3}$D, CO, GeH $_{4}$, NH $_{3}$, and PH
  $_{3}$}}.
\newblock \emph{\bibinfo{journal}{\icarus}} \textbf{\bibinfo{volume}{89}},
  \bibinfo{pages}{168--189} (\bibinfo{year}{1991}).

\bibitem{Kothari2024}
\bibinfo{author}{{Kothari}, H.} \emph{et~al.}
\newblock \bibinfo{title}{{Probing the Heights and Depths of Y Dwarf
  Atmospheres: A Retrieval Analysis of the JWST Spectral Energy Distribution of
  WISE J035934.06{\textendash}540154.6}}.
\newblock \emph{\bibinfo{journal}{\apj}} \textbf{\bibinfo{volume}{971}},
  \bibinfo{pages}{121} (\bibinfo{year}{2024}).

\bibitem{Morley18}
\bibinfo{author}{{Morley}, C.~V.} \emph{et~al.}
\newblock \bibinfo{title}{{An L Band Spectrum of the Coldest Brown Dwarf}}.
\newblock \emph{\bibinfo{journal}{\apj}} \textbf{\bibinfo{volume}{858}},
  \bibinfo{pages}{97} (\bibinfo{year}{2018}).

\bibitem{Tremblin15}
\bibinfo{author}{{Tremblin}, P.} \emph{et~al.}
\newblock \bibinfo{title}{{Fingering Convection and Cloudless Models for Cool
  Brown Dwarf Atmospheres}}.
\newblock \emph{\bibinfo{journal}{\apjl}} \textbf{\bibinfo{volume}{804}},
  \bibinfo{pages}{L17} (\bibinfo{year}{2015}).







\clearpage

\bibitem{Jakobsen2022}
\bibinfo{author}{{Jakobsen}, P.} \emph{et~al.}
\newblock \bibinfo{title}{{The Near-Infrared Spectrograph (NIRSpec) on the
  James Webb Space Telescope. I. Overview of the instrument and its
  capabilities}}.
\newblock \emph{\bibinfo{journal}{\aap}} \textbf{\bibinfo{volume}{661}},
  \bibinfo{pages}{A80} (\bibinfo{year}{2022}).

\bibitem{Glasse2015}
\bibinfo{author}{{Glasse}, A.} \emph{et~al.}
\newblock \bibinfo{title}{{The Mid-Infrared Instrument for the James Webb Space
  Telescope, IX: Predicted Sensitivity}}.
\newblock \emph{\bibinfo{journal}{\pasp}} \textbf{\bibinfo{volume}{127}},
  \bibinfo{pages}{686} (\bibinfo{year}{2015}).

\bibitem{Bushouse24}
\bibinfo{author}{Bushouse, H.} \emph{et~al.}
\newblock \bibinfo{title}{Jwst calibration pipeline} (\bibinfo{year}{2024}).
\newblock \urlprefix\url{https://doi.org/10.5281/zenodo.12692459}.

\bibitem{Greenfield16}
\bibinfo{author}{{Greenfield}, P.} \& \bibinfo{author}{{Miller}, T.}
\newblock \bibinfo{title}{{The Calibration Reference Data System}}.
\newblock \emph{\bibinfo{journal}{Astronomy and Computing}}
  \textbf{\bibinfo{volume}{16}}, \bibinfo{pages}{41--53}
  (\bibinfo{year}{2016}).

\bibitem{Gardner2006}
\bibinfo{author}{{Gardner}, J.~P.} \emph{et~al.}
\newblock \bibinfo{title}{{The James Webb Space Telescope}}.
\newblock \emph{\bibinfo{journal}{\ssr}} \textbf{\bibinfo{volume}{123}},
  \bibinfo{pages}{485--606} (\bibinfo{year}{2006}).

\bibitem{Nissen2004}
\bibinfo{author}{{Nissen}, P.~E.}
\newblock \bibinfo{editor}{{McWilliam}, A.} \& \bibinfo{editor}{{Rauch}, M.}
  (eds) \emph{\bibinfo{title}{{Thin and Thick Galactic Disks}}}.
\newblock (eds \bibinfo{editor}{{McWilliam}, A.} \& \bibinfo{editor}{{Rauch},
  M.}) \emph{\bibinfo{booktitle}{Origin and Evolution of the Elements}},
  \bibinfo{pages}{154} (\bibinfo{year}{2004}).
\newblock \eprint{astro-ph/0310326}.

\bibitem{Jofre2011}
\bibinfo{author}{{Jofr{\'e}}, P.} \& \bibinfo{author}{{Weiss}, A.}
\newblock \bibinfo{title}{{The age of the Milky Way halo stars from the Sloan
  Digital Sky Survey}}.
\newblock \emph{\bibinfo{journal}{\aap}} \textbf{\bibinfo{volume}{533}},
  \bibinfo{pages}{A59} (\bibinfo{year}{2011}).

\bibitem{Filippazzo20}
\bibinfo{author}{{Filippazzo}, J.}
\newblock \bibinfo{title}{{SEDkit: Spectral energy distribution construction
  and analysis tools}}.
\newblock \bibinfo{howpublished}{Astrophysics Source Code Library, record
  ascl:2011.014} (\bibinfo{year}{2020}).
\newblock \eprint{2011.014}.

\bibitem{Filippazzo15}
\bibinfo{author}{{Filippazzo}, J.~C.} \emph{et~al.}
\newblock \bibinfo{title}{{Fundamental Parameters and Spectral Energy
  Distributions of Young and Field Age Objects with Masses Spanning the Stellar
  to Planetary Regime}}.
\newblock \emph{\bibinfo{journal}{\apj}} \textbf{\bibinfo{volume}{810}},
  \bibinfo{pages}{158} (\bibinfo{year}{2015}).

\bibitem{Gaarn2022}
\bibinfo{author}{{Gaarn}, J.} \emph{et~al.}
\newblock \bibinfo{title}{{The puzzle of the formation of T8 dwarf Ross 458c}}.
\newblock \emph{\bibinfo{journal}{\mnras}} \textbf{\bibinfo{volume}{521}},
  \bibinfo{pages}{5761--5775} (\bibinfo{year}{2023}).

\bibitem{emcee}
\bibinfo{author}{{Foreman-Mackey}, D.}, \bibinfo{author}{{Hogg}, D.~W.},
  \bibinfo{author}{{Lang}, D.} \& \bibinfo{author}{{Goodman}, J.}
\newblock \bibinfo{title}{{emcee: The MCMC Hammer}}.
\newblock \emph{\bibinfo{journal}{\pasp}} \textbf{\bibinfo{volume}{125}},
  \bibinfo{pages}{306} (\bibinfo{year}{2013}).

\bibitem{freedman2008}
\bibinfo{author}{{Freedman}, R.~S.}, \bibinfo{author}{{Marley}, M.~S.} \&
  \bibinfo{author}{{Lodders}, K.}
\newblock \bibinfo{title}{{Line and Mean Opacities for Ultracool Dwarfs and
  Extrasolar Planets}}.
\newblock \emph{\bibinfo{journal}{\apjs}} \textbf{\bibinfo{volume}{174}},
  \bibinfo{pages}{504--513} (\bibinfo{year}{2008}).

\bibitem{freedman2014}
\bibinfo{author}{{Freedman}, R.~S.} \emph{et~al.}
\newblock \bibinfo{title}{{Gaseous Mean Opacities for Giant Planet and
  Ultracool Dwarf Atmospheres over a Range of Metallicities and Temperatures}}.
\newblock \emph{\bibinfo{journal}{\apjs}} \textbf{\bibinfo{volume}{214}},
  \bibinfo{pages}{25} (\bibinfo{year}{2014}).

\bibitem{Grimm2021}
\bibinfo{author}{{Grimm}, S.~L.} \emph{et~al.}
\newblock \bibinfo{title}{{HELIOS-K 2.0 Opacity Calculator and Open-source
  Opacity Database for Exoplanetary Atmospheres}}.
\newblock \emph{\bibinfo{journal}{\apjs}} \textbf{\bibinfo{volume}{253}},
  \bibinfo{pages}{30} (\bibinfo{year}{2021}).

\bibitem{Li2015}
\bibinfo{author}{{Li}, G.} \emph{et~al.}
\newblock \bibinfo{title}{{Rovibrational Line Lists for Nine Isotopologues of
  the CO Molecule in the X $^{1}${\ensuremath{\Sigma}}$^{+}$ Ground Electronic
  State}}.
\newblock \emph{\bibinfo{journal}{\apjs}} \textbf{\bibinfo{volume}{216}},
  \bibinfo{pages}{15} (\bibinfo{year}{2015}).

\bibitem{Rothman2010}
\bibinfo{author}{{Rothman}, L.~S.} \emph{et~al.}
\newblock \bibinfo{title}{{HITEMP, the high-temperature molecular spectroscopic
  database}}.
\newblock \emph{\bibinfo{journal}{\jqsrt}} \textbf{\bibinfo{volume}{111}},
  \bibinfo{pages}{2139--2150} (\bibinfo{year}{2010}).

\bibitem{Visscher12}
\bibinfo{author}{{Visscher}, C.}
\newblock \bibinfo{title}{{Chemical Timescales in the Atmospheres of Highly
  Eccentric Exoplanets}}.
\newblock \emph{\bibinfo{journal}{\apj}} \textbf{\bibinfo{volume}{757}},
  \bibinfo{pages}{5} (\bibinfo{year}{2012}).

\bibitem{kass1995}
\bibinfo{author}{Kass, R.~E.} \& \bibinfo{author}{Raftery, A.~E.}
\newblock \bibinfo{title}{Bayes factors}.
\newblock \emph{\bibinfo{journal}{Journal of the American Statistical
  Association}} \textbf{\bibinfo{volume}{90}}, \bibinfo{pages}{773--795}
  (\bibinfo{year}{1995}).
\newblock
  \urlprefix\url{http://www.tandfonline.com/doi/abs/10.1080/01621459.1995.10476572}.

\bibitem{Gonzales2021}
\bibinfo{author}{{Gonzales}, E.~C.} \emph{et~al.}
\newblock \bibinfo{title}{{The First Retrieval of a Substellar Subdwarf: A
  Cloud-free SDSS J125637.13-022452.4}}.
\newblock \emph{\bibinfo{journal}{\apj}} \textbf{\bibinfo{volume}{923}},
  \bibinfo{pages}{19} (\bibinfo{year}{2021}).

\bibitem{Piskorz2018}
\bibinfo{author}{{Piskorz}, D.} \emph{et~al.}
\newblock \bibinfo{title}{{Ground- and Space-based Detection of the Thermal
  Emission Spectrum of the Transiting Hot Jupiter KELT-2Ab}}.
\newblock \emph{\bibinfo{journal}{\aj}} \textbf{\bibinfo{volume}{156}},
  \bibinfo{pages}{133} (\bibinfo{year}{2018}).

\bibitem{Arcangeli2018}
\bibinfo{author}{{Arcangeli}, J.} \emph{et~al.}
\newblock \bibinfo{title}{{H$^{-}$ Opacity and Water Dissociation in the
  Dayside Atmosphere of the Very Hot Gas Giant WASP-18b}}.
\newblock \emph{\bibinfo{journal}{\apjl}} \textbf{\bibinfo{volume}{855}},
  \bibinfo{pages}{L30} (\bibinfo{year}{2018}).

\bibitem{Gharib-Nezhad2019}
\bibinfo{author}{{Gharib-Nezhad}, E.} \& \bibinfo{author}{{Line}, M.~R.}
\newblock \bibinfo{title}{{The Influence of H$_{2}$O Pressure Broadening in
  High-metallicity Exoplanet Atmospheres}}.
\newblock \emph{\bibinfo{journal}{\apj}} \textbf{\bibinfo{volume}{872}},
  \bibinfo{pages}{27} (\bibinfo{year}{2019}).

\bibitem{Welbanks2024}
\bibinfo{author}{{Welbanks}, L.} \emph{et~al.}
\newblock \bibinfo{title}{{A high internal heat flux and large core in a warm
  Neptune exoplanet}}.
\newblock \emph{\bibinfo{journal}{\nat}} \textbf{\bibinfo{volume}{630}},
  \bibinfo{pages}{836--840} (\bibinfo{year}{2024}).

\bibitem{Kirkpatrick21}
\bibinfo{author}{{Kirkpatrick}, J.~D.} \emph{et~al.}
\newblock \bibinfo{title}{{The Field Substellar Mass Function Based on the
  Full-sky 20 pc Census of 525 L, T, and Y Dwarfs}}.
\newblock \emph{\bibinfo{journal}{\apjs}} \textbf{\bibinfo{volume}{253}},
  \bibinfo{pages}{7} (\bibinfo{year}{2021}).

\bibitem{Zhang2017}
\bibinfo{author}{{Zhang}, Z.~H.} \emph{et~al.}
\newblock \bibinfo{title}{{Primeval very low-mass stars and brown dwarfs - II.
  The most metal-poor substellar object}}.
\newblock \emph{\bibinfo{journal}{\mnras}} \textbf{\bibinfo{volume}{468}},
  \bibinfo{pages}{261--271} (\bibinfo{year}{2017}).

\bibitem{Lodders2002}
\bibinfo{author}{{Lodders}, K.} \& \bibinfo{author}{{Fegley}, B.}
\newblock \bibinfo{title}{{Atmospheric Chemistry in Giant Planets, Brown
  Dwarfs, and Low-Mass Dwarf Stars. I. Carbon, Nitrogen, and Oxygen}}.
\newblock \emph{\bibinfo{journal}{\icarus}} \textbf{\bibinfo{volume}{155}},
  \bibinfo{pages}{393--424} (\bibinfo{year}{2002}).

\bibitem{Zachariah93}
\bibinfo{author}{{Zachariah}, M.~R.} \& \bibinfo{author}{{Tsang}, W.}
\newblock \bibinfo{title}{{Application ofAb InitioMolecular Orbital and
  Reaction Rate Theories to Nucleation Kinetics}}.
\newblock \emph{\bibinfo{journal}{Aerosol Science Technology}}
  \textbf{\bibinfo{volume}{19}}, \bibinfo{pages}{499--513}
  (\bibinfo{year}{1993}).

\bibitem{Zachariah95}
\bibinfo{author}{Zachariah, M.~R.} \& \bibinfo{author}{Tsang, W.}
\newblock \bibinfo{title}{Theoretical calculation of thermochemistry,
  energetics, and kinetics of high-temperature sixhyoz reactions}.
\newblock \emph{\bibinfo{journal}{The Journal of Physical Chemistry}}
  \textbf{\bibinfo{volume}{99}}, \bibinfo{pages}{5308--5318}
  (\bibinfo{year}{1995}).
\newblock \urlprefix\url{https://doi.org/10.1021/j100015a012}.

\bibitem{Meisner23RN}
\bibinfo{author}{{Meisner}, A.~M.} \emph{et~al.}
\newblock \bibinfo{title}{{Deep DECam Y-band Follow-up of WISEA
  J153429.75-104303.3 (a.k.a. ``The Accident'')}}.
\newblock \emph{\bibinfo{journal}{Research Notes of the American Astronomical
  Society}} \textbf{\bibinfo{volume}{7}}, \bibinfo{pages}{36}
  (\bibinfo{year}{2023}).

\bibitem{Meisner23}
\bibinfo{author}{{Meisner}, A.~M.} \emph{et~al.}
\newblock \bibinfo{title}{{Exploring the Extremes: Characterizing a New
  Population of Old and Cold Brown Dwarfs}}.
\newblock \emph{\bibinfo{journal}{\aj}} \textbf{\bibinfo{volume}{166}},
  \bibinfo{pages}{57} (\bibinfo{year}{2023}).

\bibitem{Marocco21}
\bibinfo{author}{{Marocco}, F.} \emph{et~al.}
\newblock \bibinfo{title}{{The CatWISE2020 Catalog}}.
\newblock \emph{\bibinfo{journal}{\apjs}} \textbf{\bibinfo{volume}{253}},
  \bibinfo{pages}{8} (\bibinfo{year}{2021}).

\bibitem{Brewer16}
\bibinfo{author}{{Brewer}, J.~M.} \& \bibinfo{author}{{Fischer}, D.~A.}
\newblock \bibinfo{title}{{C/O and Mg/Si Ratios of Stars in the Solar
  Neighborhood}}.
\newblock \emph{\bibinfo{journal}{\apj}} \textbf{\bibinfo{volume}{831}},
  \bibinfo{pages}{20} (\bibinfo{year}{2016}).

\bibitem{Akerman2004}
\bibinfo{author}{{Akerman}, C.~J.}, \bibinfo{author}{{Carigi}, L.},
  \bibinfo{author}{{Nissen}, P.~E.}, \bibinfo{author}{{Pettini}, M.} \&
  \bibinfo{author}{{Asplund}, M.}
\newblock \bibinfo{title}{{The evolution of the C/O ratio in metal-poor halo
  stars}}.
\newblock \emph{\bibinfo{journal}{\aap}} \textbf{\bibinfo{volume}{414}},
  \bibinfo{pages}{931--942} (\bibinfo{year}{2004}).

\end{thebibliography}
\end{document}